\documentclass[structabstract]{aa}  
\usepackage{graphicx}
\usepackage{txfonts}
\usepackage{natbib}
\bibpunct{(}{)}{;}{a}{}{,}
\def\gapprox{{_>\atop{^\sim}}}
\def\lapprox{{_<\atop{^\sim}}}

\def\cmmd{\rm {cm^{-3}}}
\def\cmmt{\rm {cm^{-2}}}

\def\s-1{\rm {s^{-1}}}

\def\HC3N{HC$_3$N}

\def\kms{\hbox{${\rm km\,s}^{-1}$}}
\def\msun{M$_{\odot}$}
\def\lsun{L$_{\odot}$}

\usepackage{url}

\begin{document}
 \title{High resolution observations of HCN and HCO$^+$ $J$=3--2 in the disk and outflow of Mrk~231\thanks{Based on observations
carried out with the IRAM Plateau de Bure Interferometer. IRAM is supported by INSU/CNRS (France), MPG (Germany) and IGN (Spain)}}

   \subtitle{Detection of vibrationally excited HCN in the warped nucleus.}

   \author{S. Aalto
          \inst{1}
	  \and
	  S. Garcia-Burillo\inst{2}
          \and
          S. Muller\inst{1} 
          \and
          J. M. Winters\inst{3}
          \and
          E. Gonzalez-Alfonso\inst{4}
	  \and	
          P. van der Werf\inst{5}
          \and
          C. Henkel\inst{7,8}
          \and
	  F. Costagliola\inst{6,1}
          \and
	  R. Neri\inst{3}
             }

 \institute{Department of Earth and Space Sciences, Chalmers University of Technology, Onsala Observatory,
              SE-439 94 Onsala, Sweden\\
              \email{saalto@chalmers.se}
       \and Observatorio Astron\'omico Nacional (OAN)-Observatorio de Madrid, Alfonso XII 3, 28014-Madrid, Spain
       \and Institut de Radio Astronomie Millim\'etrique (IRAM), 300 rue de la Piscine, Domaine Universitaire de Grenoble,
38406 St. Martin d$ ' $H\`eres, France
       \and Universidad de Alcal\'a de Henares,Departamento de F\'isica y Matem\'aticas, Campus Universitario, 28871 Alcal\'a de Henares, Madrid, Spain 
       \and Leiden Observatory, Leiden University, 2300 RA, Leiden, The Netherlands
       \and Instituto de Astrof\'isica de Andaluc\'ia, Glorieta de la Astronom\'ia, s/n, 18008, Granada, Spain
       \and Max-Planck-Institut f{\"u}r Radioastronomie, Auf dem H{\"u}gel 69, 53121 Bonn, Germany
 \and Astronomy Department, King Abdulaziz University, P.O. Box 80203
    Jeddah 21589, Saudi Arabia
	  }

   \date{Received xx; accepted xx}

 
  \abstract
   {}
   {Our goal is to study molecular gas properties in nuclei and large scale outflows/winds from active galactic nuclei and starburst galaxies.}
   {We obtained high resolution (0.\arcsec 25 to 0.\arcsec 90) observations of HCN and HCO$^+$ $J=3\to 2$ of the ultraluminous 
   QSO galaxy Mrk~231 with the IRAM Plateau de Bure Interferometer.}
   {We find luminous HCN and HCO$^+$ $J=3\to 2$ emission in the main disk and we detect compact ($r \lapprox$ 0.\arcsec 1 (90 pc)) vibrationally excited HCN $J=3\to 2$ $\nu_2$=1$f$ emission centred on the nucleus.  The velocity field of the vibrationally excited HCN is strongly inclined (position angle PA=155$^{\circ}$) compared to the east-west rotation of the main disk. The nuclear ($r \lapprox$ 0.\arcsec 1) molecular mass is estimated to $8 \times10^8$ \msun\ with an average $N$(H$_2$) of $1.2 \times10^{24}$ $\cmmt$.  
Prominent, spatially extended ($\gapprox$350 pc) line wings are found for HCN $J=3\to 2$ with velocities up to $\pm 750$ \kms.
Line ratios indicate that the emission is emerging in dense gas $n=10^4 - 5 \times 10^5$ $\cmmd$ of elevated HCN abundance $X$(HCN)=$10^{-8} - 10^{-6}$. High $X$(HCN) also allows for the emission to originate in gas of more moderate density.  We tentatively detect nuclear emission from the reactive ion HOC$^+$ with HCO$^+$/HOC$^+$=10--20.
}
   {The HCN $\nu_2$=1$f$ line emission is consistent with the notion of a hot, dusty, warped inner disk of Mrk~231 where the $\nu_2$=1$f$ line is excited by bright mid-IR 14 $\mu$m continuum. We estimate the vibrational temperature $T_{\rm vib}$ to 200-400 K.
 Based on relative source sizes we propose that 50\% of the main HCN emission may have its excitation affected by the radiation field through IR pumping of the vibrational ground state.  The HCN emission in the line wings, however, is more extended and thus likely not strongly affected by IR pumping. Our results reveal that dense clouds survive (and/or are formed) in the AGN outflow on scales of at least several hundred pc before evaporating or collapsing. The elevated HCN abundance in the outflow is consistent with warm chemistry 
possibly related to shocks and/or X-ray irradiated gas. An upper limit to the mass and momentum flux is $4 \times 10^8$ \msun\ and 12$L_{\rm AGN}/c$, respectively, and we discuss possible driving mechanisms for the dense outflow.
   }
    \keywords{galaxies: evolution
--- galaxies: individual: Mrk~231
--- galaxies: active
--- quasars: general
--- ISM: jets and outflows
--- ISM: molecules
               }

  \maketitle


\section{Introduction}

The ultraluminous infrared galaxy (ULIRG, log($L_{\rm IR})=12.37$) Mrk~231
is often referred to as the closest IR quasar (QSO). It has been identified as a major merger, and it hosts powerful AGN activity
as well as a young, dusty starburst with an extreme star formation rate of $\approx$ 200 \msun
yr$^{-1}$ \citep{taylor99,gallagher02,lipari09}. Large amounts of molecular
gas are located in an almost face-on east-west rotating disk \citep{bryant,downes98}.
Mrk~231 shows evidence of various types of mechanical feedback from the nuclear activity including
radio jets (from pc to kpc scales) \citep{carilli98,lonsdale03} and optical absorption lines tracing starburst driven
winds and the jet acceleration of ionized and atomic gas to high velocities ($\approx$ 1400 \kms) \citep{rupke11}. Mrk~231 also has a
powerful molecular outflow manifested by broad (750 \kms) CO wings \citep{feruglio10,cicone12} and in OH-absorption by \citet{fischer10}
and recently at very high velocities (1500 \kms) in highly excited OH \citep{gonzalez14}. \citet{feruglio10} estimate the outflow of molecular
gas as 700 \msun \,yr$^{-1}$, emptying the reservoir of cold gas within 10~Myrs. \\

The high mass outflow rates seem to support the notion that the bulk of the gas is driven out by the AGN \citep[e.g.][]{feruglio10,rupke11}. 
However, little is known about the properties of the molecular gas in the outflow. Reliable estimates of molecular masses will rely on determining
the physical conditions of the molecular gas in the outflow. The properties of the gas and dust within the inner few hundred pc of Mrk~231 are still poorly
constrained, including how they relate to the nuclear activity and the molecular gas in the outflow. \\

Recently, we have found that in the Mrk~231 outflow the HCN/CO $J=1\to 0$ line ratio is very
high with a ratio 0.3 -- 1 \citep{aalto12a}. This is even higher than in the line core where the HCN luminosity is already 
unusually high \citep{solomon92}. The HCN emission covers the same velocity width
($\pm$ 750 \kms) as the one seen in CO 1--0 by \citep{feruglio10}.  We suggested that the high brightness of HCN 1--0 is a result of
the molecular mass in the outflow residing mostly in dense ($n \gapprox 10^4$ $\cmmd$) gas.
However, the possibility that the luminous HCN emission was due to  mid-IR pumping via the 14 $\mu$m bending modes could not be entirely dismissed. 
In particular, the CO emission seems to be subthermally excited in the outflow \citep{cicone12}, indicating low to moderate densities.
Recently, rotational transitions of HCN within the vibrationally excited state $\nu_2$=1 have been detected towards the LIRG NGC~4418 \citep{sakamoto10}
and towards the ULIRG IRAS~20551-4250 \citep{imanishi13}. Radiative excitation may populate the $\nu_2$=1 ladder and may in turn also pump the $\nu$=0
vibrational ground state levels affecting the excitation of the whole molecule. It is therefore fundamental to investigate the excitation
of molecular emission towards luminous IR galaxies so that the molecular properties in both nuclei and outflows may be understood.\\

\noindent
In the paper we present IRAM Plateau de Bure A- and B-array HCN, HCO$^+$ $J=3\to 2$ high resolution data of the disk and outflow of Mrk~231.
We also detected the $J=3\to 2$ $\nu_2$=1$f$ vibrational line and use it to further understand the properties
and kinematics of the very nuclear gas.  From now on we refer to the HCN and HCO$^+$ $J=3\to 2$ lines in the vibrational
ground state ($\nu$=0) as HCN and HCO$^+$ 3--2 (apart from Sec~\ref{s:vib-temp} where we use $\nu_2$=1/$\nu$=0 for the line ratio between 
HCN $J=3\to 2$ $\nu_2$=1$f$ and $\nu$=0). The HCN $J=3\to 2$ $\nu_2$=1$f$ vibrational line will be referred to as HCN 3--2 $\nu_2$=1$f$.


\section{Observations}
\label{s:obs}

The 1mm observations were carried out with the six-element IRAM Plateau de Bure array in 
February 2012 in A-array and in February 2013 in B-array.  The phase centre was set to $\alpha$=12:56:14.232 and $\delta$= 56:52:25.207 (J2000).
Mrk~231 was observed for hour angles 2 to 8 and the radio seeing ranged between $0.\arcsec 22$ and $0.\arcsec 48$.
The receivers were tuned to a frequency of 255.88~GHz, the centre between the two red-shifted line frequencies  for HCN ($\nu$=255.13~GHz)
and HCO$^+$ ($\nu$=256.73~GHz). We used the WideX correlator providing a broad frequency range of 3.6~GHz.

The bandpass of the individual antennas was derived from the bright quasar 3C273.
The flux calibration was set on 3C273.  The quasars close to Mrk~231 $J1150+497$ ($\sim 2$ Jy at 1 mm) and
$J1418+546$ ($\sim 0.55$ Jy) were observed regularly for complex gain calibration every 25 minutes.

After calibration within the GILDAS reduction package, the visibility
set was converted into FITS format, and imported into the GILDAS/MAPPING and AIPS packages
for further imaging.  We merged the A and B array data sets, rebinning to 40~MHz (47 \kms) cleaned with a natural weighting resulting 
in a beam size of 0.\arcsec 39 $\times$  0.\arcsec 35 (100 mJy=14 K) and position angle PA=0$^{\circ}$ (rms noise is 0.8~mJy~channel$^{-1}$). 
By default it is this combined dataset that we use when we discuss the results in this paper. However,  we also use the A- and B-array datasets
individually where the A-array data were cleaned with a robust weighting of -2, resulting 
in a beam size of 0.\arcsec28 $\times$  0.\arcsec25 (100 mJy=25 K) and position angle PA=59$^{\circ}$ (rms noise is 1.6~mJy~channel$^{-1}$).
The B-array data were tapered and cleaned with a natural weighting resulting 
in a beam size of 0.\arcsec91 $\times$  0.\arcsec81 (100 mJy=2.4 K) and position angle PA=83$^{\circ}$ (rms noise is 1.5~mJy~channel$^{-1}$). 

Zero velocity refers to the redshift $z$=0.042170 \citep{carilli98} throughout the paper, resulting in a spatial scale of 0.\arcsec 1=87~pc.


\section{Results}
\label{s:res}
Line fluxes, source size fits (FWHM), line widths ($\Delta$V) and emission position angles (PA), including
fits to the line wing emission, are presented in Tab.~\ref{t:flux}. 

\subsection{HCN 3--2}

\subsubsection{Integrated intensities}

\begin{figure*}
\resizebox{18cm}{!}{\includegraphics[angle=0]{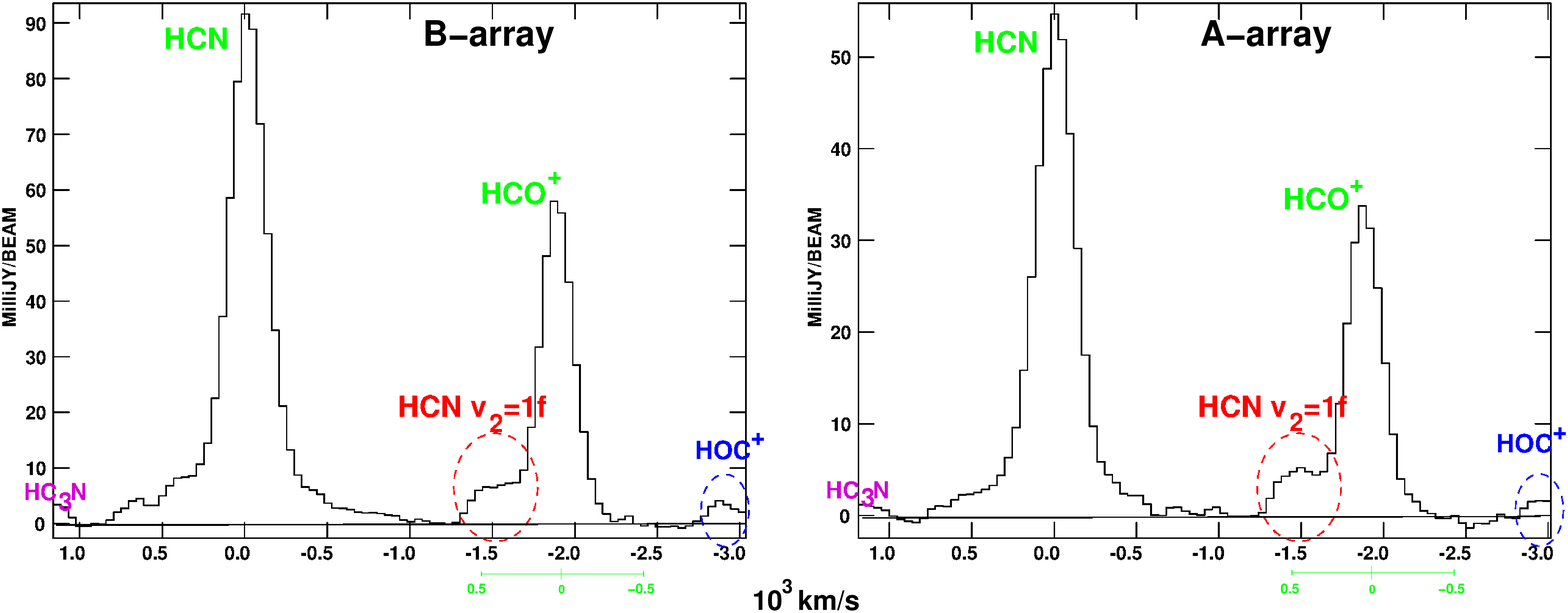}}
\caption{\label{f:spec} Separate A- and B-array spectra towards the central beam. Spectra have been Gaussian smoothed (over two channels). We show the spectra
for the A- and B-array separately to emphasize that identified features appear in both separate datasets. The central spectrum for the combined A+B dataset is shown in Fig.~\ref{f:gauss}.
{\it Left:} B-array spectrum of HCN $J$=3--2, HCO$^+$ $J$=3--2, HCN $J$=3--2 $\nu_2$=1, and a
tentative detection of HOC$^+$ $J$=3--2. {\it Right:}  A-array spectrum of the same lines. The green velocity scale is for HCO$^+$.
}
\end{figure*}

The HCN 3--2 emission consists of luminous emission from the  {\it line core} ($\pm 250$ \kms) and higher velocity, fainter {\it line wing} emission (Fig.~\ref{f:spec}). We divide the 
presentation of the HCN 3--2 integrated intensities into two sections - one where we discuss the line core emission emerging from the main disk and one where we discuss the line wing emission:\\

\begin{figure*}
\resizebox{6cm}{!}{\includegraphics[angle=0]{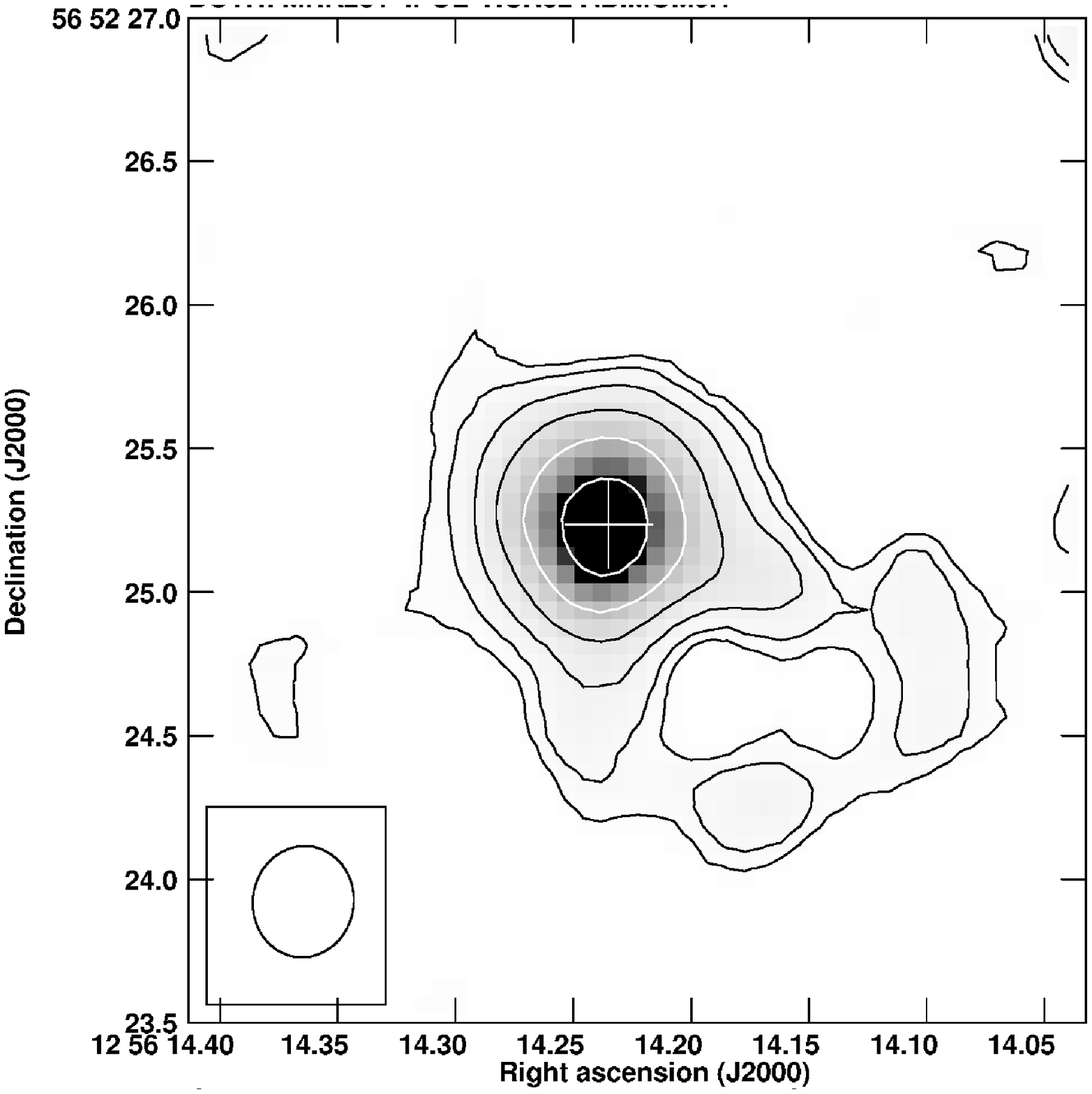}}
\resizebox{6cm}{!}{\includegraphics[angle=0]{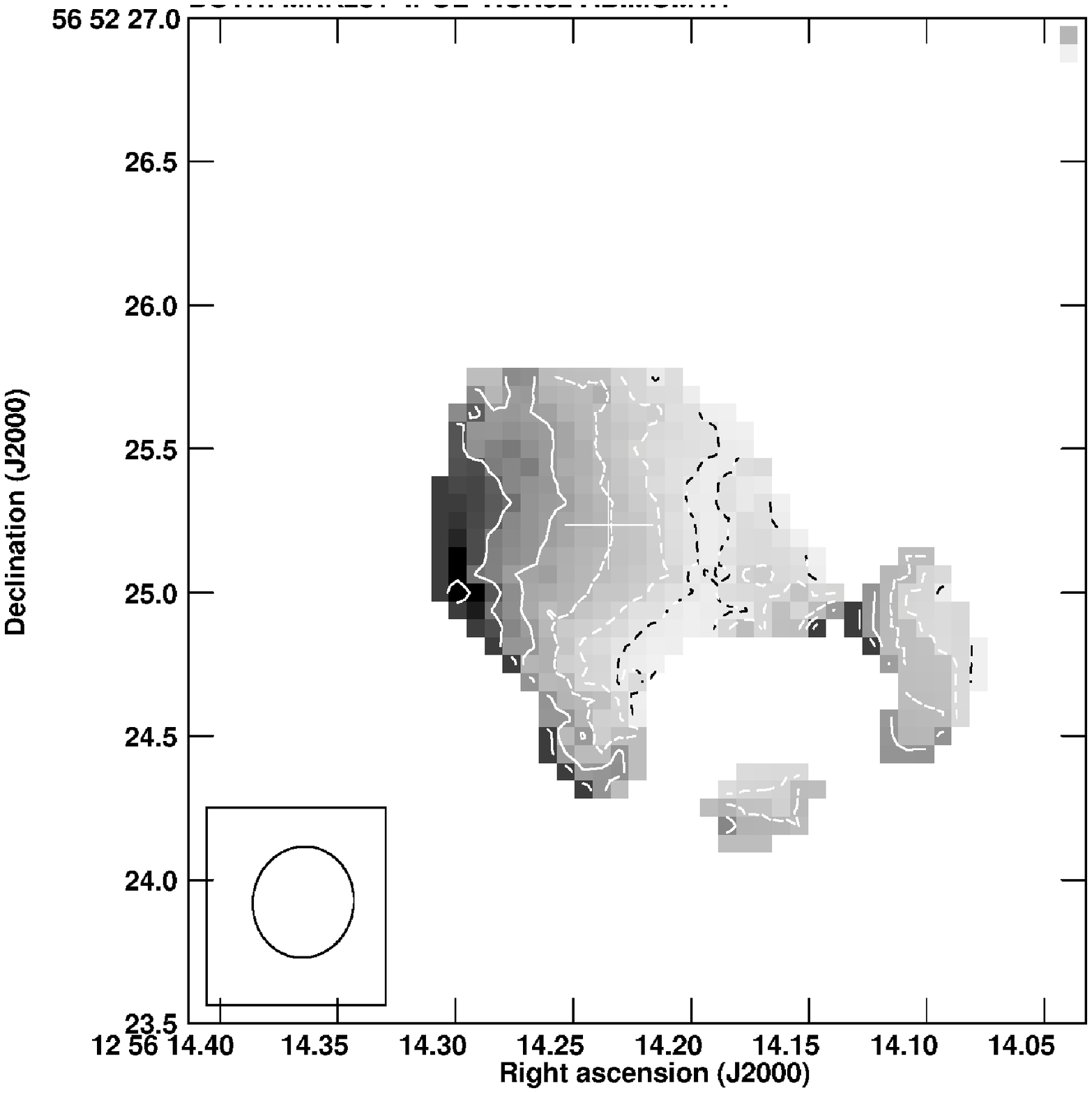}}
\resizebox{6cm}{!}{\includegraphics[angle=0]{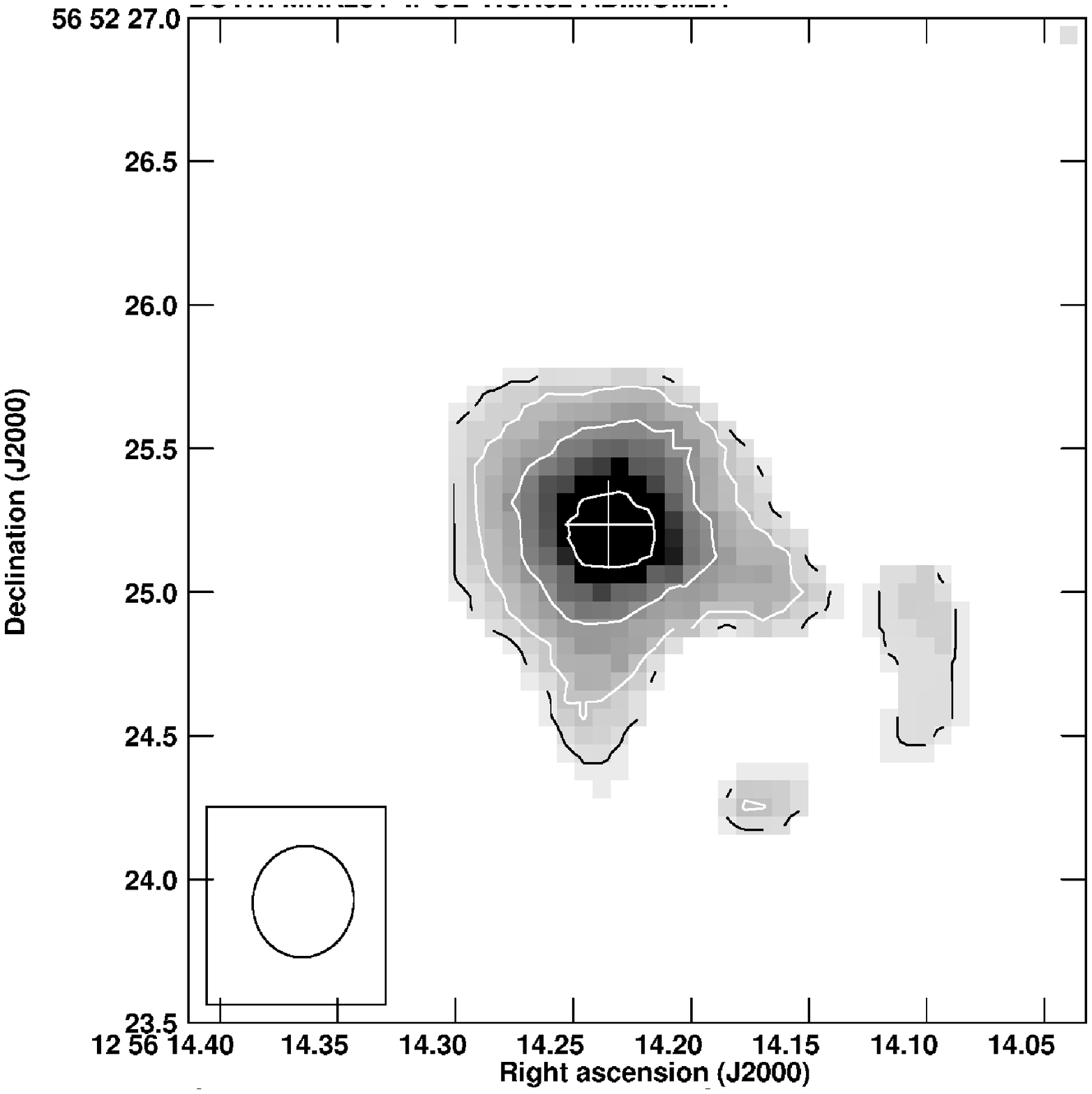}}
\resizebox{6cm}{!}{\includegraphics[angle=0]{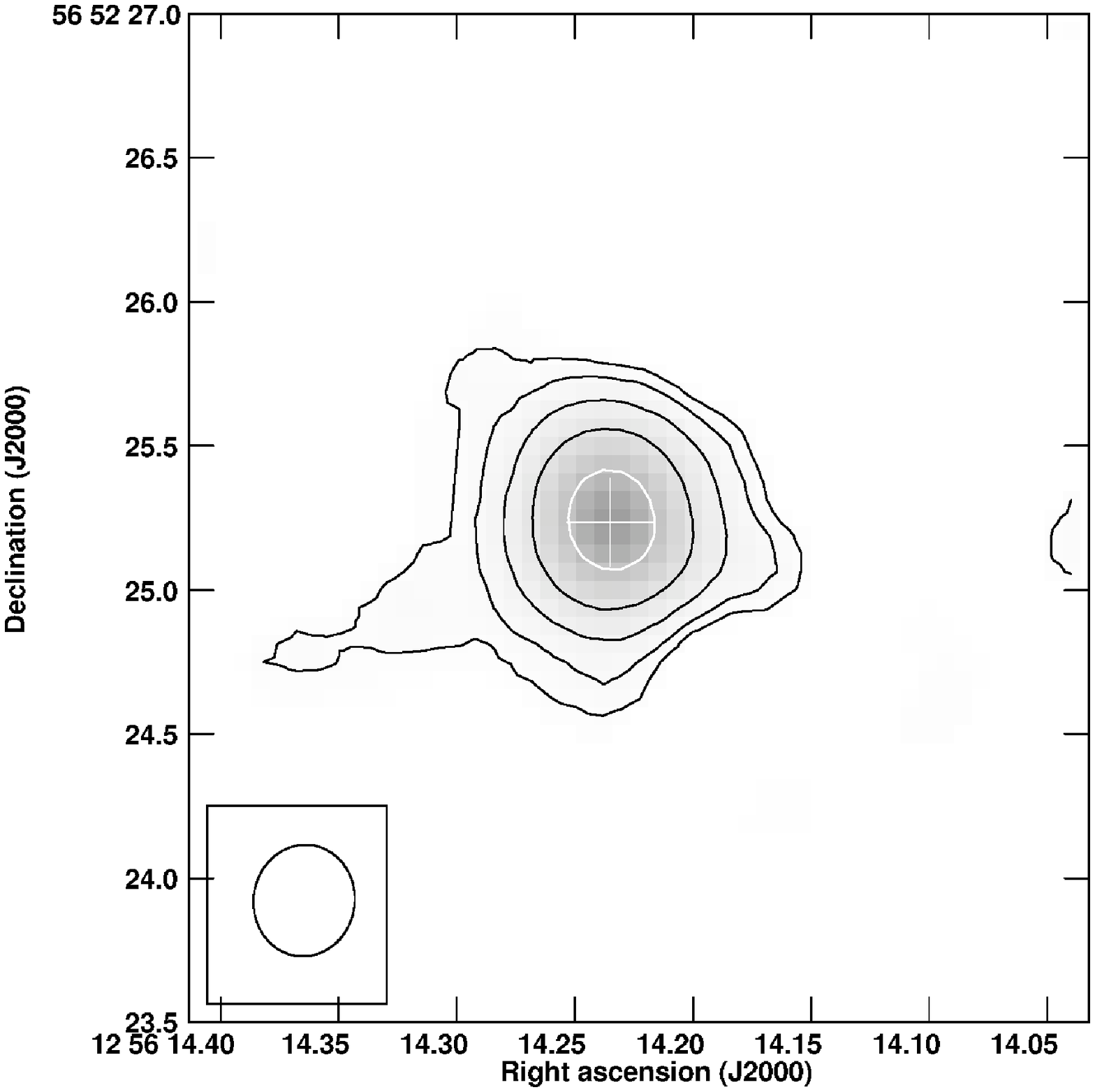}}
\resizebox{6cm}{!}{\includegraphics[angle=0]{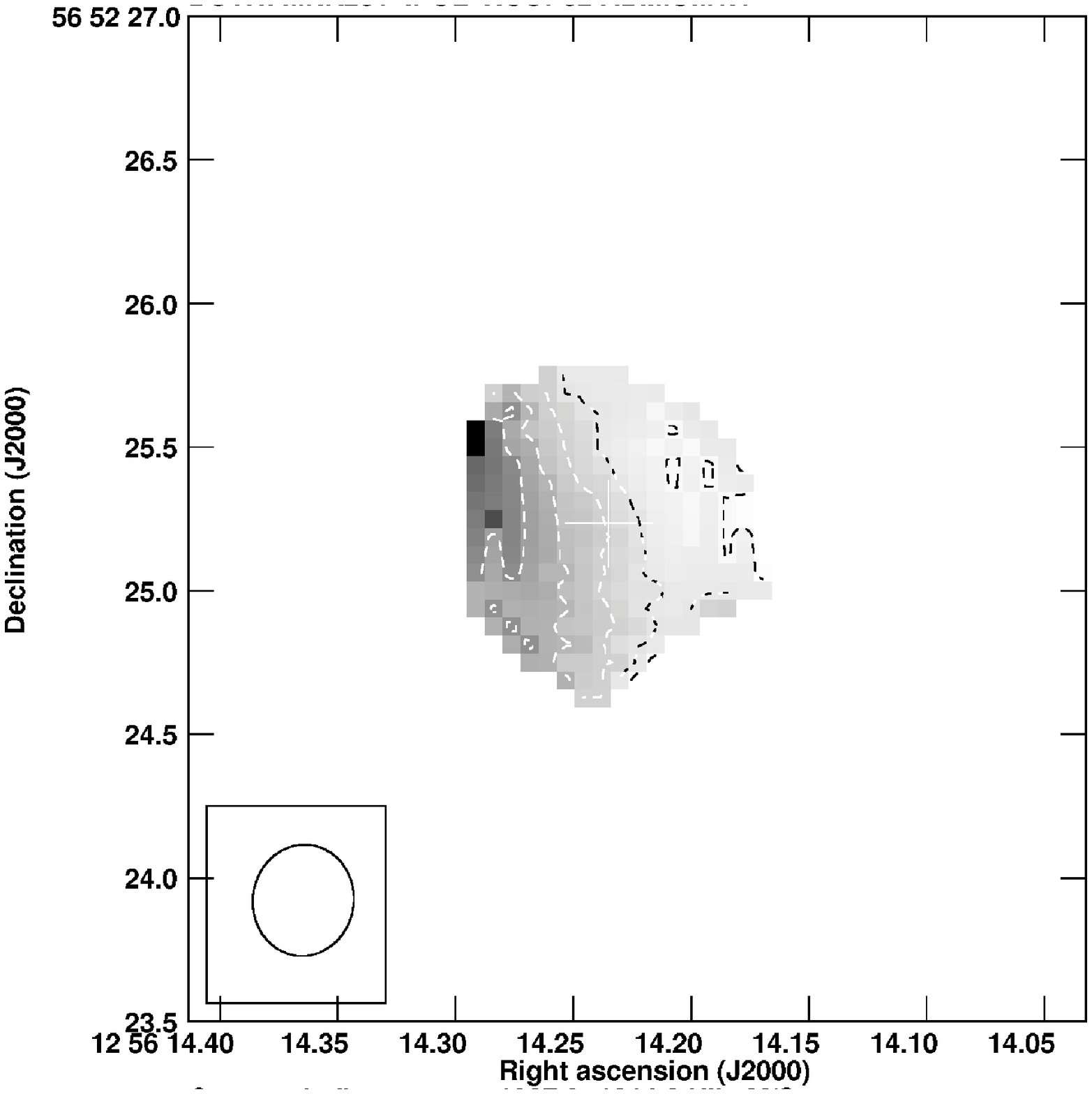}}
\resizebox{6cm}{!}{\includegraphics[angle=0]{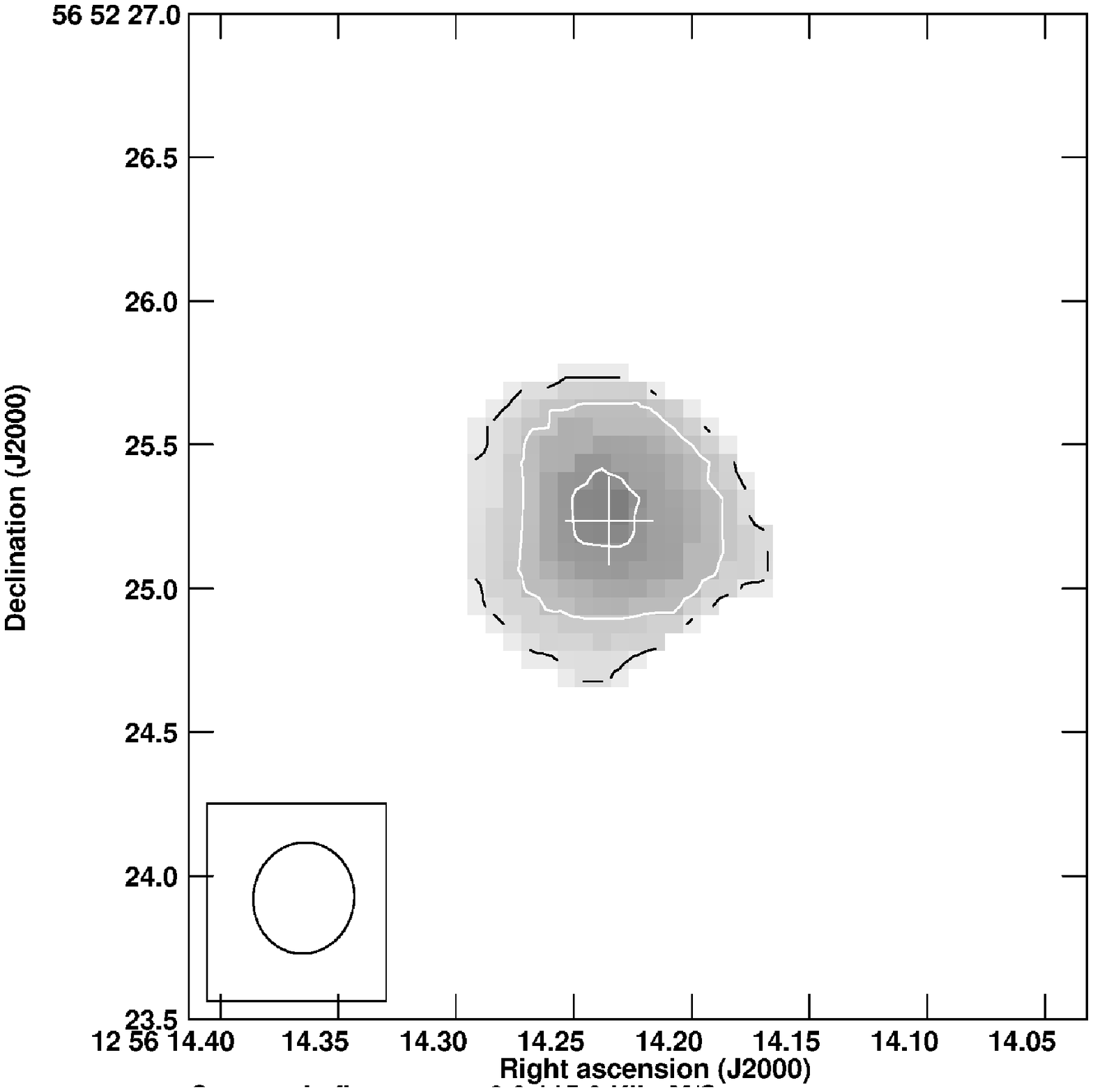}}
\caption{\label{f:mom} {\bf Top panels:} Combined A+B moment maps of the HCN 3--2 line emission from the main disk (as defined in the text). 
{\it Left panel:} Integrated intensity with contour levels 
$0.5 \times (1,2,4,8,16,32)$ Jy~beam$^{-1}$~\kms\ and grey (colour) scale ranging from 0 to 15 Jy~beam$^{-1}$~\kms\ (peak
integrated flux is 22 Jy~beam$^{-1}$~\kms).  {\it Centre:}  Velocity field with contour levels starting at -56~\kms\ and then increasing
by steps of 18.8~\kms\ until +38~\kms. The grey (colour) scale ranges from -65 to 50~\kms. {\it Right:} Dispersion map where contours
start at 30~\kms\ and then increase
by steps of 30~\kms. Greyscale ranges from 0 to 115~\kms\  and peak dispersion is 130~\kms. The cross marks the position
of the peak HCN 3--2 integrated intensity. 
 {\bf Lower panels:} Combined A+B moment maps of the HCO$^+$ 3--2 line. {\it Left panel:} Integrated intensity with contour levels 
$0.5 \times (1,2,4,8,16,32)$ Jy~beam$^{-1}$~\kms\ and grey (colour) scale ranging from 0 to 15 Jy~beam$^{-1}$~\kms\ (peak
integrated flux is 12 Jy~beam$^{-1}$~\kms).  {\it Centre:}  Velocity field with contour levels starting at -56~\kms\ and then increasing
by steps of 18.8~\kms\ until +38~\kms. The grey (colour) scale ranges from -65 to 50~\kms. {\it Right:} Dispersion map where contours
start at 30~\kms\ and then increase
by steps of 30~\kms. Greyscale ranges from 0 to 115~\kms\  and peak dispersion is 98~\kms. The cross marks the position
of the peak HCN 3--2 integrated intensity. 
}
\end{figure*}

\noindent
{\it Main disk:} \, The HCN 3--2 emission is distributed in a bright nuclear component surrounded by extended disk emission (Fig.~\ref{f:mom}).
The lower-surface brightness emission extends out to a radius of at least 0.\arcsec8 (700~pc). 
\citet{downes98} use CO 2--1 position-velocity (pV) maps to estimate the disk size along the line of nodes indicating a 0.\arcsec6 radius inner disk
and a 1.\arcsec5 outer disk - where the line of nodes has a  PA=90$^{\circ}$. The HCN 3--2 emission thus appears to be residing in the inner part of the CO disk.

Tongues of fainter emission extend at least 1\arcsec (870 pc) from the nucleus of Mrk~231 and they connect to the south-west in what appears like a ring-like 
feature.  It is possible that this structure is the result of a remnant side-lobe (and then would not be real). However, side-lobes should be symmetric and there is
no evidence of a similar feature on the other side of the nucleus. The feature is also brighter than the expected side-lobe response.
Furthermore, Mrk~231 shows evidence of distinct optical circular shell like features at a whole range of radii - resulting from a complicated pattern of
outflows and winds \citep{lipari05}. Our HCN 3--2 ring-like feature coincides with their shell feature S3. \\

\noindent
{\it Line wings:} \, HCN 3--2 line wings are detected out to velocities  of at least 750 \kms\ (Fig.~\ref{f:spec}), similar to those found in HCN 1--0 by \citet{aalto12a}. 
We integrated the line wings over velocities $V$=350 -- 990 \kms\ to be safely away from the rotation ($V_{\rm rot}$ $\pm$ 150 \kms) of the main disk of
Mrk~231 and to be consistent with the fits to the HCN 1--0 line wings.

\begin{figure*}
\resizebox{16cm}{!}{\includegraphics[angle=0]{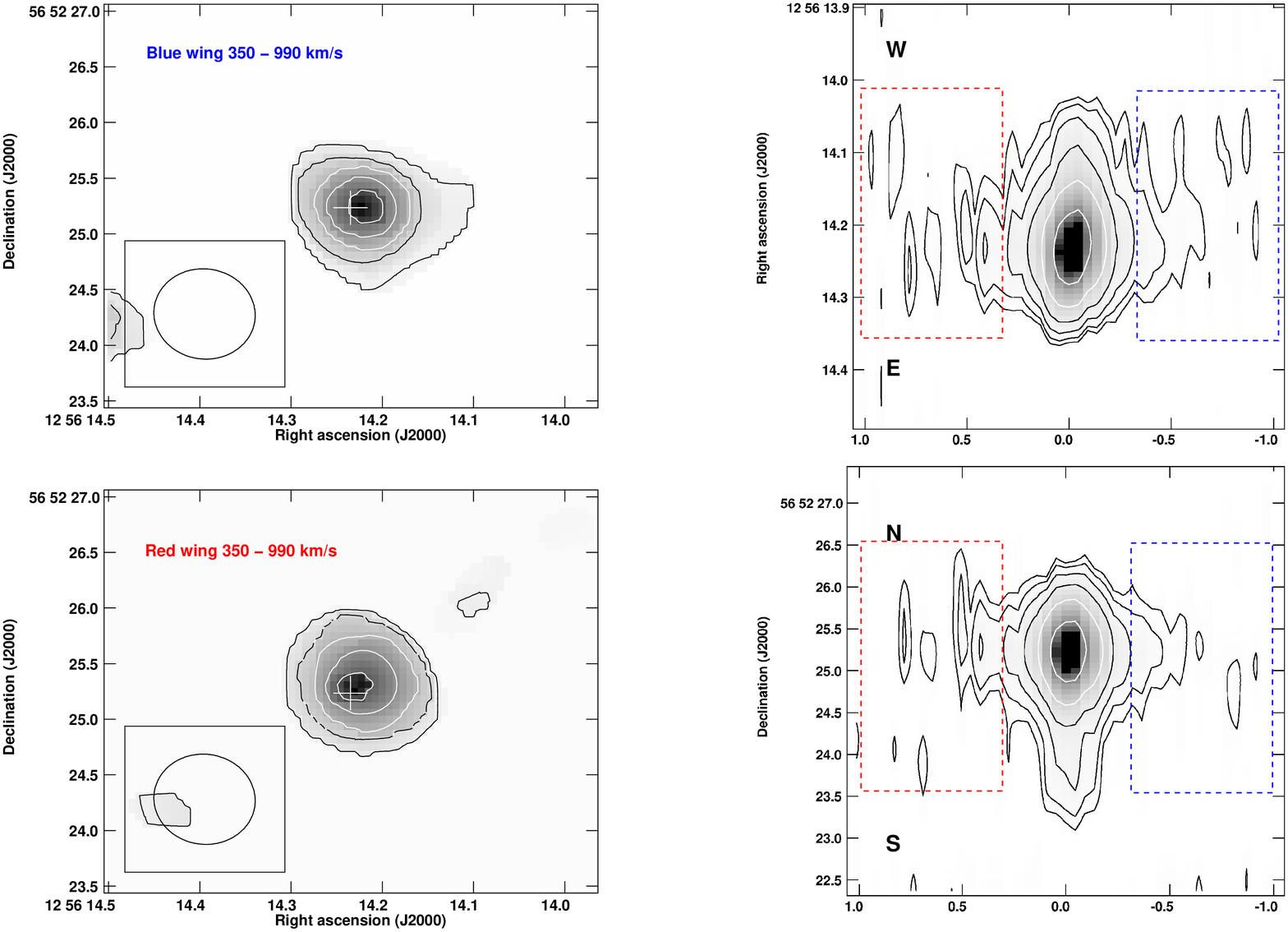}}
\caption{\label{f:wings} {\it Left:} Moment 0 maps (integrated intensity) of the HCN 3--2 line wings in tapered B array.  Contour levels
are $0.2 \times (1,3,5,7,9)$ Jy~beam$^{-1}$~\kms\ and greyscale ranges from 0 to 2 Jy~beam$^{-1}$~\kms. The cross marks the position
of the peak HCN 3--2 integrated intensity. 
{\it Right:} pV diagrams of the tapered B array map. {\bf Upper:} East-west cut along the line of nodes 
of the main disk rotation. {\bf Lower:} North-south cut along the minor axis of the main disk rotation.
Greyscale ranges from 0 to 120 mJy beam$^{-1}$, contours are 3$\times$(1,2,4,8,16,32)  mJy beam$^{-1}$.
First contour is 2$\sigma$
}
\end{figure*}

The line wings are faint and spatially extended. In the A+B combined and naturally weighted integrated intensity map  (with 3$\sigma$
cut-off) we recover only 50\% of the flux in the wings compared to the tapered (0.\arcsec91 $\times$  0.\arcsec81) B-array data. This is
despite the lower sensitivity in the B-array only data.  Even when we taper the combined A+B array data to 0.\arcsec9 resolution we still
miss 20\% of the flux recovered in the B-array data. Therefore, we use the B-array data to study the extent and total flux of the HCN 3--2 line wings.
A two-dimensional Gaussian fit to the integrated intensity maps results in FWHM sizes of 0.\arcsec4 - 0.\arcsec 6 (350 - 520 pc)(see Fig.~\ref{f:wings}). 
The spatial shift between the red- and blue wings is smaller than the beam: the blue wing is shifted 0.\arcsec 15 to the west of the nucleus
and the red wing 0.\arcsec 1 to the north. 

Since the line wings are broad and relatively faint, the determination of their positions depends on a reliable continuum subtraction.
We subtracted the continuum in the uv plane and tried both a fit to the line-free spectral channels of order 0 and 
order 1 (linear fit allowing a slope). We then compared the results of the two methods and conclude that there is no measurable difference
between them.


\begin{table*}
\caption{\label{t:flux} Line flux densities$^a$.}
\begin{tabular}{lcccccccc}
\hline
\hline\\[0.01 pt]
Line & $f_{\rm rest}$ & $E_{\rm u}/k$ & Peak & $\Delta V$ & Integrated & Luminosity$^b$ & Source size & PA  \\
\hline \\[0.01ex]
& [GHz] & [K] & [mJy\,beam$^{-1}$] & [\kms] & [Jy \kms] & [$10^7$ K \kms pc$^{-2}$] & [FWHM] & [$^{\circ}$] \\[0.01ex]
\hline \\ 
{\bf HCN 3--2} & 265.886 & 25.5 & 21.5 $\pm$ 0.2 & 308 & 38.2 $\pm$ 0.4  & 36.7 & 0.\arcsec33 $\times$ 0.\arcsec33 ($\pm$0.\arcsec01)   & $50 \pm 10$ \\
{\it Red wing}$^b$ &&& 1.9 $\pm$ 0.2 & $\dots$ & 2.5 $\pm$ 0.3  & 2.4 & $>$0.\arcsec4   & $\dots$ \\
{\it Blue wing}$^c$ &&& 1.2 $\pm$ 0.1 & $\dots$ & 1.9 $\pm$ 0.1  &1.8 & $>$0.\arcsec4    & $\dots$ \\
\\
{\bf HCN 3--2 $\nu_2$=1$f$} & 267.199 & 1050.0 & 1.5 $\pm$ 0.1 & $\dots$ & 1.6 $\pm$ 0.2 & 1.5 & 0.\arcsec17 $\times$ 0.\arcsec16 ($\pm$0.\arcsec01)  & $155 \pm 10$\\
\\
{\bf HCO$^+$ 3--2} & 267.558  & 25.7 & 10.7 $\pm$ 0.1 & 235 & 20.1 $\pm$ 0.1 & 19.3 & 0.\arcsec37 $\times$ 0.\arcsec36 ($\pm0."01$) & $25 \pm 5$ \\ 
\\
{\bf HOC$^+$ 3--2 } & 268.451 & 25.6 &  0.6 $\pm$ 0.1 & $\dots$ & 0.7 $\pm$ 0.1 & $\dots$ & $\dots$ & $\dots$\\
\\
\hline \\
\end{tabular} 

\noindent
a) Fluxes are calculated from two-dimensional Gaussian fits to the integrated intensity maps of the A+B combined dataset (apart from the line wings - see b) below). Maps
were produced from integrating flux above the 4$\sigma$ cut-off. The integrated intensity (for all lines) peaks at  $\alpha$:12:56:14.237,   $\delta$: 56:52:25.23 (J2000).
Positional errors from the fitting are lower than 0.\arcsec 002 (rms). Adding effects of phase and calibration errors the positional accuracy is roughly 0.\arcsec 02. \\
\noindent
b) The line luminosity is $L$=$\pi R^2 I \approx \pi R^2T_{\rm B}1.06\Delta V$). \\
\noindent
c) From two-dimensional Gaussian fits to the integrated intensity maps tapered to 0.\arcsec9 resolution in the B-array data. The line wings are integrated between
$\pm$(350 and 990) \kms\ and for the luminosity we assume that all the 
flux is contained inside the tapered B-array beam (where 100 mJy=2.4 K).

\end{table*}


\subsubsection{Velocity field,  channel map and pV diagrams}

\begin{figure}
\resizebox{8cm}{!}{\includegraphics[angle=0]{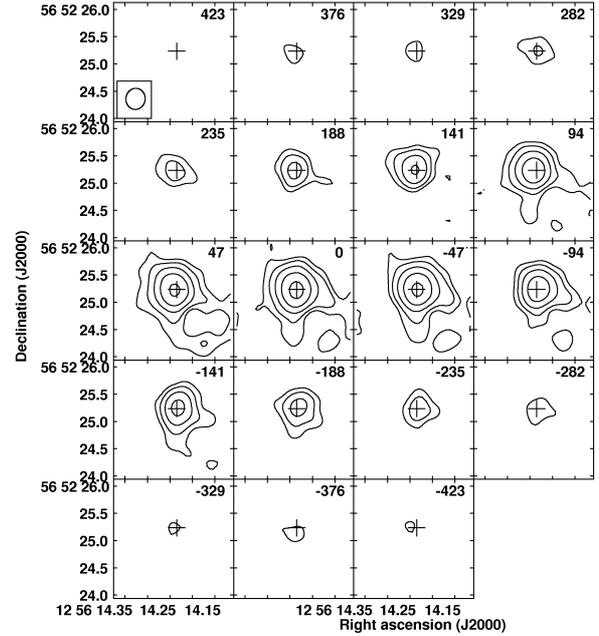}}
\caption{\label{f:chan} Channel map of the HCN 3--2 emission towards Mrk~231,
as obtained from the combination of the A and B
configuration data. Contours are logarithmic: 3.6$\times$(1,2,4,8,16) mJy~beam$^{-1}$ 
where the first contour is at 4.5$\sigma$.
The synthesized beam of 0.\arcsec 39 $\times$  0.\arcsec 35  is shown in the upper left panel.}
\end{figure}

The HCN 3--2 velocity field and dispersion map are presented in Fig.~\ref{f:mom}. 
The velocity field indicates rotation with the line of nodes at PA=90$^{\circ}$ tracing the east-west rotation of the main disk- just as the CO 2--1 \citep{downes98}.
However, the channel map in Fig.~\ref{f:chan} suggests a more complex velocity structure and the isovelocity contours in the velocity field clearly twist away from 
a simple spider diagram of orderly east-west rotation. The dispersion map peaks on the nucleus - but there is also a suggestion
that the dispersion is tracing extentions to the south, north and west.

\begin{figure}
\resizebox{8cm}{!}{\includegraphics[angle=0]{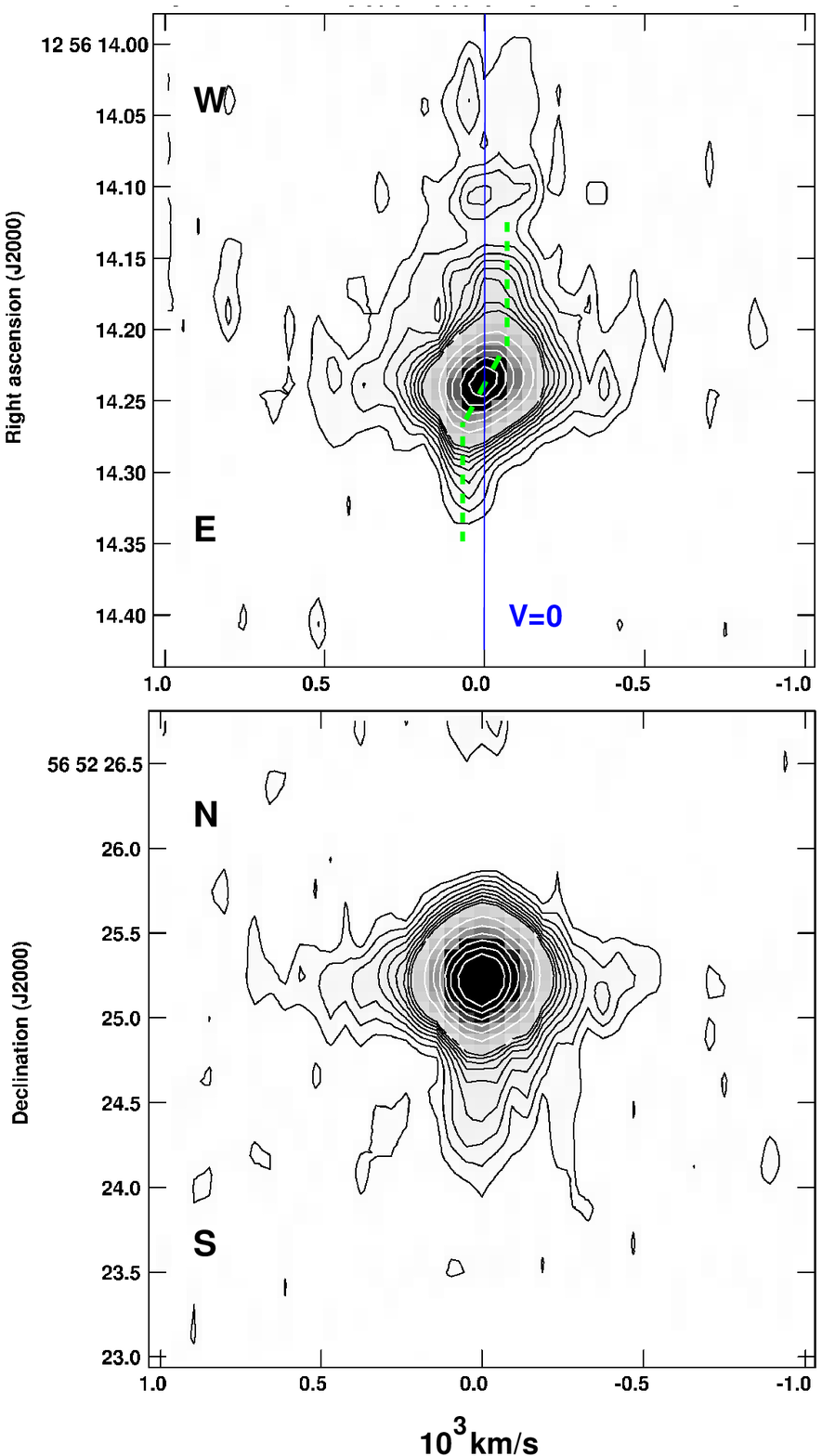}}
\caption{\label{f:pv} HCN 3--2 position-velocity (pV) diagrams of the A+B combined data: {\it Top: }  East-west cut along the line of nodes 
of the main disk rotation.  {\it Lower: } North-south cut along the minor axis of the main disk rotation.
Greyscale ranges from 0 to 40 mJy beam$^{-1}$, contours are 1.4$\times$(1,2,3,4,5,6,7,8,9,15,20,25,30) mJy beam$^{-1}$. First contour is 2$\sigma$.
 (1\arcsec=870~pc) . Green dashed line outline the stellar velocity curve from \citet{davies}.
}
\end{figure}

\noindent
{\it Position-velocity (pV) diagrams} \, HCN 3--2 pV diagrams are presented in Fig.~\ref{f:pv}.  We have cut along the east-west axis (along the line of nodes of the main disk
rotation) and the north-south axis. In the east-west cut the dominant rotation of the main disk is evident. Emission close to systemic velocity is found  1.\arcsec 5 (1.3 kpc)
to the south and 2.\arcsec 5 (2.2 kpc) to the west of the emission peak. These features are found also in the HCN 1--0 map \citep{aalto12a}. The western feature is found at
forbidden velocities, redder than $V$=0 by 50-100 \kms, which means that they do not represent a simple extension of the rotating main disk. 

In Fig.~\ref{f:wings} we present the tapered B-array data pV diagrams to better show the emission from the line wings. Their extended nature is evident as is the fact that
the spatial shift between the red and blue wings is smaller than their extent.


\subsection{Vibrational HCN 3--2 $\nu_2$=1$f$}
\label{s:res_vib}

\begin{figure}
\resizebox{8cm}{!}{\includegraphics[angle=0]{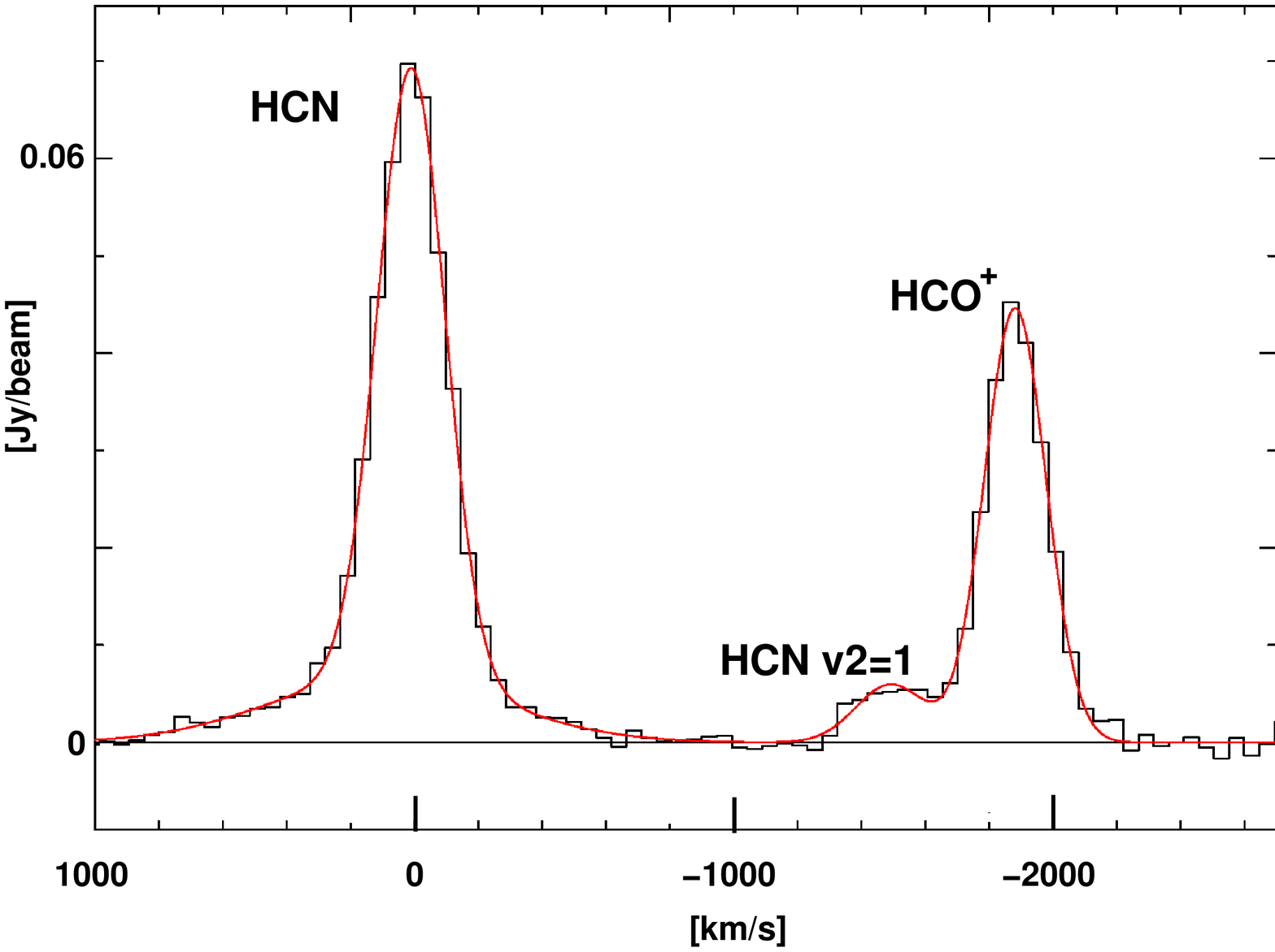}}
\caption{\label{f:gauss} Gaussian fits to the A+B combined data central spectrum. Fits are peak flux and line width: HCN $\nu$=0 line centre (63 mJy/beam, 250 \kms), 
line wings (7 mJy/beam, 840 \kms); HCN $\nu_2$=1$f$ (6 mJy/beam, 260 \kms); HCO$^+$ (45 mJy/beam, 227 \kms). 
Since the A- and B-array data had a slight shift in centre frequency the edges of the band have been cut, leaving out the HOC$^+$ detection
at the blue end of the spectrum.
}
\end{figure}

For the first time the HCN J=3--2 $\nu_2$=1$f$ line is detected in Mrk~231 (see Fig.~\ref{f:spec} for A and B-array spectra, Fig~\ref{f:gauss} for A+B combined array spectrum). 
The integrated intensity shows a compact unresolved structure centred on the nucleus (Fig.~\ref{f:vib}). The velocity field of the
HCN 3--2 $\nu_2$=1$f$ line (Fig.~\ref{f:vib}) has a different position angle than that of the HCN 3--2 main line (Fig.~\ref{f:mom}). Its line of nodes are 
oriented with a PA of 150$^{\circ} \pm 20^{\circ}$.  A Gaussian fit to the HCN and HCO$^+$ lines in the combined A+B spectra  (see Fig.~\ref{f:gauss}) suggest  a line width (FWHM) of
257 \kms\  ($\pm$ 10 \kms) for the HCN 3--2 $\nu_2$=1$f$ line.  Note that the HCN J=3--2 $\nu_2$=1$f$ line is
located 400 \kms\ to the red side of the HCO$^+$ 3--2 line which may cause some line blending when lines are broad.

\subsubsection{Potential alternative interpretations}

There are two possible alternative interpretations of the emission feature that we identify as the vibrational HCN 3--2 $\nu_2$=1$f$ line. One is that it is a wing to the HCO$^+$ 3--2 line and the 
other is that it is the SO 4$_3$ -- 3$_4$ line. Below we discuss these possibilities and why we dismiss them.\\ 

{\it Why is this not a wing of the HCO$^+$ line?} \, The HCN 3--2 $\nu_2$=1$f$ line is expected to have its maximum intensity 400 \kms\ to the red side of the HCO$^+$ line
which is where we find it.  Furthermore, the  HCN 3--2 $\nu_2$=1$f$ line emission is compact and the source size fit is $\lapprox$ 0.\arcsec 2 compared to $>$0.\arcsec 4 for the HCN line wings.
In addition, the HCN 3--2 $\nu_2$=1$f$ line has a distinct kinematic pattern (see above) that is different from that of the HCN wind  in the vibrational ground state.

\noindent
{\it Could this instead be an SO line? } \, It is important to consider the possibility that the feature identified as the HCN 3--2 $\nu_2$=1$f$ line could
instead be the SO 4$_3$ -- 3$_4$ line. Since we have both 2mm and 3mm interferometric data of Mrk~231 we can test for this possibility
through searching for other SO lines. We ran RADEX non-LTE radiative transport models \citep{vandertak}  for a range of physical
conditions ($T_{\rm k}$=20 - 200 K; $n=10^3 - 10^6$ $\cmmd$) and we find that the 2mm 178~GHz 4$_5$ -- 3$_4$ SO line is always two orders of magnitude brighter than
the 4$_3$ -- 3$_4$ line. Fortunately the 178~GHz SO line falls into the band of our HCN 2--1 PdBI data (Lindberg et. al. in prep.). We do not see any signs of
this line and put a very conservative upper limit to its brightness of of 4 mJy. The corresponding 4$_3$ -- 3$_4$ line would not be detectable
in our data.  Thus we conclude that we have correctly identified the feature 400 \kms\ redwise of HCO$^+$ 3--2 as the HCN 3--2 $\nu_2$=1$f$ line.


\subsection{HCO$^+$ 3--2}
\label{s:hco}

For HCO$^+$ 3--2 we detect only the line core emission - no line wings. The emission feature 400 \kms\ redshifted from HCO$^+$ 3--2 is identified as 
the HCN 3--2 $\nu_2$=1$f$ vibrational line (see previous section).

The HCO$^+$ 3--2 integrated intensity map of the total line emission is presented in Fig.~\ref{f:mom}.  The size and structure of the map
is similar to that of the HCN 3--2 even if the total flux is lower by a factor of 1.5 - 2.  The HCO$^+$ 3--2 emission has a slightly larger
source size than HCN 3--2 due to the absence of strongly peaked nuclear emission. Part of the intensity difference between HCN and HCO$^+$ is
also due to the broader HCN line.

The velocity field and dispersion map are presented in Fig.~\ref{f:mom}. The velocity dispersion is clearly lower than that of HCN 3--2
and the difference in line FWHM is more than 70 \kms\ (see also Tab.~\ref{f:mom}) The velocity field is dominated by
the east-west rotation already seen in the HCN 3--2 map.

\subsection{HOC$^+$ $J$=3--2}
\label{s:hco}

At the edge of the band (Fig.~\ref{f:spec}) we tentatively detect the HOC$^+$ $J$=3--2  line ($\nu$=268.451 GHz and redshifted to 257.588 GHz). We do not pick up the complete
HCO$^+$ line so we can only estimate the HCO$^+$/HOC$^+$ 3--2 line ratio to 10--20 based on the peak emission. For comparison the HCO$^+$/HOC$^+$ 1--0
ratio for the nearby Seyfert galaxy NGC~1068 is 80 (integrated, 40--50 peak temperature ratio)\citep{usero04} implying that the HOC$^+$ line emission is relatively brighter in
Mrk~231. The HOC$^+$ line emission will be discussed further in Lindberg et al (in prep.).


\subsection{Continuum}
\label{s:res_cont}

Gaussian fits to the continuum in the A-array and A+B-array observations are presented in Table~\ref{t:c_flux}. If we consider the A and B- array dataset separately we find that 
in the B-array data the integrated flux density is  $44.9 \pm  1.5$ mJy, while in the A-array data the flux is a factor of 1.9 lower.  From uv fitting we concluded (before merging the datasets)
that this difference is caused by the shorter baselines of the B-array data and that a significant fraction of the 1mm continuum is extended beyond the compact nucleus and emerges
from the main disk.


\begin{figure}
\resizebox{8cm}{!}{\includegraphics[angle=0]{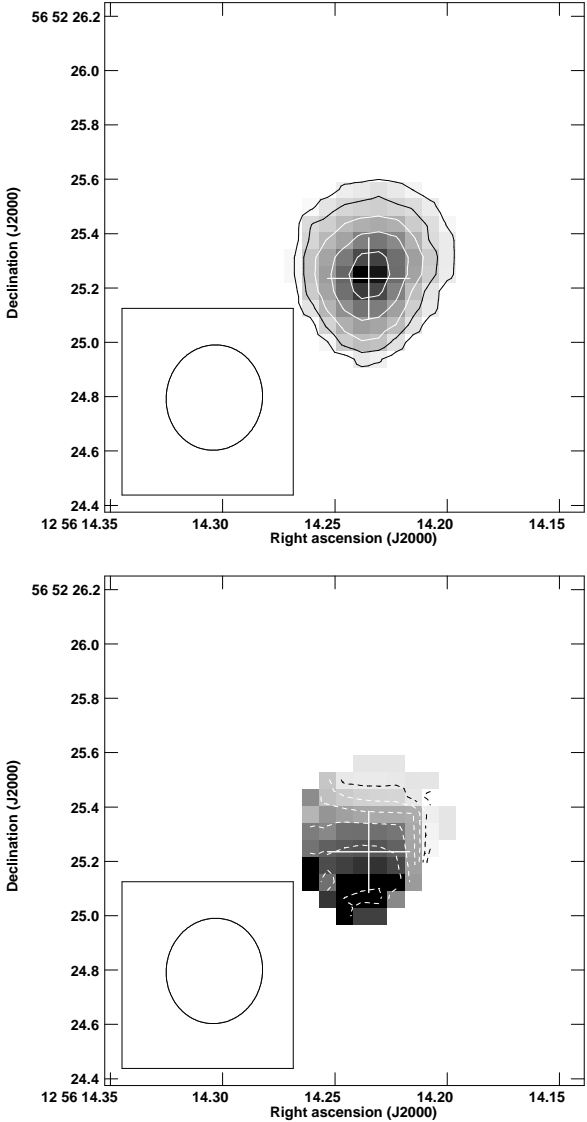}}
\caption{\label{f:vib} {\it Top panel:} Integrated intensity of the HCN J=3--2 $\nu_2$=1$f$ line emission where the contour levels 
are $0.14 \times (1,3,5,7,9)$ Jy~beam$^{-1}$~\kms. The peak flux is 1.5 Jy~beam$^{-1}$~\kms\ and the cross marks the position
of the peak HCN 3--2 integrated intensity. {\it Lower panel:} Velocity field with contour levels starting at -60~\kms\ and then increasing
by steps of 15\kms\ until +30~\kms\ . The grey (colour) scale ranges from -60 to 40~\kms. 
}
\end{figure}



\begin{table}
\caption{\label{t:c_flux} Continuum flux - 1mm (256 GHz)$^a$}
\begin{tabular}{lcccc}
\hline
\hline\\[0.01 pt]
Array & Peak & Integrated & size & PA\\
\hline \\[0.001ex]
& [mJy\,beam$^{-1}$] & [mJy] & [\arcsec] & [$^{\circ}$] \\
\hline \\ 
A & $19 \pm  1$ & $24 \pm  2$ & 0.\arcsec 22 $\times$ 0.\arcsec 02 & $7 \pm 5$ \\
A+B & $25 \pm  0.5$ & $38 \pm  1$  & 0.\arcsec 28 $\times$ 0.\arcsec 26 &$6 \pm 20$ \\
A+B$^b$ & $30 \pm  1$ & $44 \pm  2$  & 0.\arcsec 40 $\times$ 0.\arcsec 36 & $80 \pm 30$\\
\\
\hline \\
\end{tabular} 

a) From two-dimensional Gaussian fits to the integrated intensity maps. The position of the 1mm continuum source
is $\alpha$=12:56:14.234 and $\delta$=56:52:25.248 ( the total positional accuracy is
0.\arcsec 02 see footnote of Tab.~\ref{t:flux}) (from the A-array fits - the B-array position agrees within the errors). 
The Merlin 1.6 GHz continuum peak position \citep{richards05} (at similar resolution as ours) agrees with our 256~GHz continuum
position within the errors. The position is also consistent with the astrometric position at 8.4 GHz \citep{patnaik92} and with the 15 GHz
VLBA continuum position by \citet{ulvestad99}.\\
b) Tapered to the same resolution as the B-array configuration.

\end{table}



\begin{figure}
\resizebox{8cm}{!}{\includegraphics[angle=0]{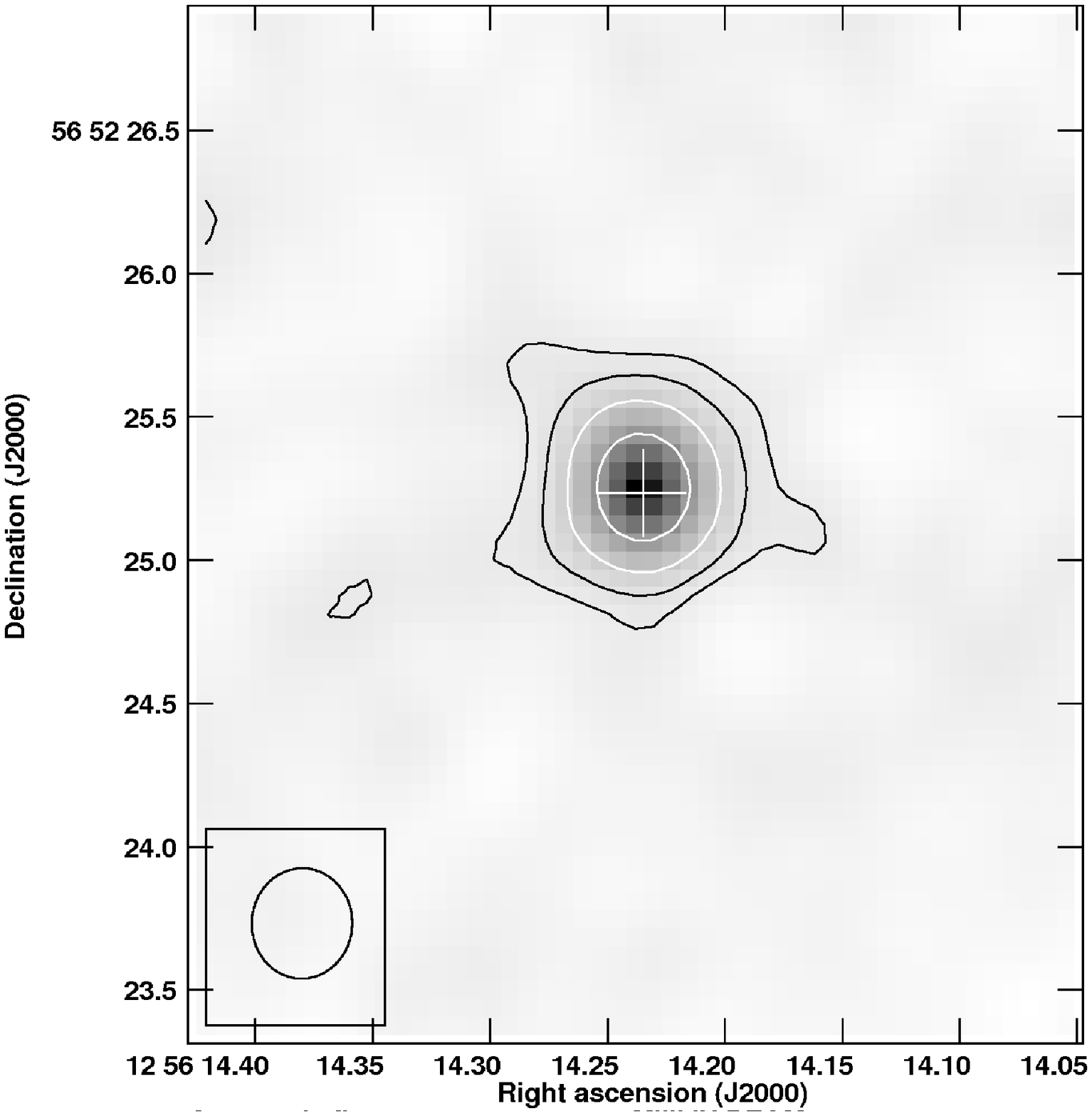}}
\caption{\label{f:cont} Integrated intensity of the 256~GHz continuum with contour levels 
$2 \times (1,2,4,8)$ mJy~beam$^{-1}$~\kms\ and grey (colour) scale ranging from -2.6 to 25 mJy~beam$^{-1}$~\kms\ (peak
integrated flux is 25 mJy~beam$^{-1}$~\kms).
}
\end{figure}

\section{Discussion}

\subsection{Vibrationally excited HCN in the QSO nucleus}
\label{s:vib}

The $J$=3--2 v$_2$=1, $l$=1$f$ line is a rotational transition within a vibrationally excited state.
The first excited bending state of HCN is doubly degenerate, so that each rotational level $\nu_2$=1 $J$ is split into a closely spaced pair of levels
($f$ and $e$). The $\nu_2$=1$e$ 3--2 line is blended with the HCN 3--2 line in the vibrational ground state ($\nu$=0) and cannot be studied for extragalactic sources
while the $\nu_2$=1$f$ line has rest frequency $\nu$=267.1993 GHz, shifted by 358 MHz (+400 \kms) from the HCO$^+$ 3--2 line. (The HCN $\nu_2$=1 $l$=1$e$ line
is right on top of the main $\nu$=0 line and will not affect the interpretation of the line shape of the main line.  It may be absorbed by the $\nu$=0 line or may boost 
the line centre emission by a similar intensity as the HCN $\nu_2$=1$f$ line.)
The energy above ground of the $\nu_2$=1 state is $T_{E/k}$=1050 K and intense $T_{\rm B}$(IR)$>$ 100 K mid-IR emission is required to
excite these states. 
The critical density of the line is large enough that collisional excitation is unlikely \citep{ziurys86,mills13}. 
HCN vibrational lines therefore directly probe the size, structure and dynamics of optically thick mid-IR dust cores.

The fits to the source size suggest that the emission is located within 0.\arcsec 2 ($r<$ 90~pc) of the AGN (Tab~\ref{t:flux}, Fig~\ref{f:vib}). This is consistent with mid-IR imaging of
Mrk~231 which reveal a compact  ($\lapprox$0.\arcsec 13) source of $T_{\rm B}$(IR)$>$141 K \citep{soifer} and with the dust model for Mrk~231 by \citet{gonzalez10}
where the mid-IR is dominated by a hot component with $T_{\rm d}$=150-400 K emerging from the inner 40 - 50~pc. Furthermore, the 14$\mu$m absorption
line ($\nu_2$=0--1) of HCN has been detected towards Mrk~231 with the {\it Spitzer} telecope \citep{lahuis07}. This shows that HCN in Mrk~231 absorbs mid-IR continuum which
can populate the vibrational $\nu_2$=1 ladder.

\subsubsection{The vibrational temperature $T_{\rm vib}$}
\label{s:vib-temp}

We can estimate a $T_{\rm vib}$ through comparing the intensities of the $\nu_2$=1$f$ and vibrational ground state ($\nu$=0) lines. We use the peak brightness ratio (=10)
from the A-array data since the $\nu_2$=1$f$ line is nuclear and compact. Assuming optically thin emission  
for both lines, and that the lines are co-spatial, we estimate $T_{\rm vib}$ to 400 K.  

\noindent
{\it If the lines are optically thick:} \, We know from the low HCN/H$^{13}$CN 1--0 ratio of 7 \citep{costagliola11}
that HCN is unlikely to be optically thin.  If we adopt an optical depth for the HCN main line $\tau$(HCN 3--2) of 10 (assuming the $\nu_2$=1$f$ line remains 
thin) $T_{\rm vib}$ is reduced to 210 K, for $\tau$(HCN 3--2)=20 $T_{\rm vib}$=185 K. We know that a firm lower limit to $T_{\rm vib}$ must be 100 K
since at lower IR  brightness temperatures the $\nu_2$=1$f$ line is not excited.  However, the $\nu_2$=1/$\nu$=0 brightness ratio drops quickly with temperature
so at our sensitivity we would not detect $T_{\rm B}({\rm vib})$ much below 200 K.

\noindent
{\it If lines are not co-spatial} \, The $\nu_2$=1$f$ emission region seems to be smaller than that of the $\nu$=0 line even in the A-array beam. 
This causes us to underestimate $T_{\rm vib}$ for the HCN $\nu_2$=1$f$  emitting region since the real
$\nu_2$=1/$\nu$=0 ratio will be higher.  

\noindent
To resolve both issues we must observe lines of multiple transitions at even higher resolution,  but we can
conclude that $T_{\rm vib} \gapprox$200 K and may exceed 400 K.

\subsubsection{Is the vibrational HCN tracing a warped disk?}
\label{s:vib_kin}

The velocity field of the HCN $J$=3--2 $\nu_2$=1$f$ line is tilted with respect to the east-west rotation of the almost face-on main disk. From stellar absorption
line studies \citet{davies} propose that the inner 0\arcsec2 - 0\arcsec3 (170-250 pc) of Mrk~231 is warped out of its original disk plane. They present
a model of the warp where they tilt the disk slightly (10$^{\circ}$) within $r<$0.\arcsec 2 and then let it decline uniformly to zero at $r>$0.\arcsec 35. The result is a
major axis of the inner part of the disk of  PA=-25$^{\circ}$. High resolution OH megamaser observations by \citep{klockner03,richards05} suggest an even stronger
effect of warping in the centre of Mrk~231 with a dusty disk or torus inclined by 56$^{\circ}$. Interestingly the OH velocity field of \citet{klockner03} is similar
to the one we find for the HCN $J$=3--2 v$_2$=1, $l$=1$f$ line (see Fig.~\ref{f:vib}).  

\citet{davies} comment that it is difficult to warp a stellar disk but point out that the stars in the disk
of Mrk~231 are young and that the warp likely occured first in a gaseous disk and that the young stars then were formed in situ.

The question is then why we do not see evidence of this warp in the inner part of the HCN 3--2 velocity structure. An inspection of the velocity field (Fig.~\ref{f:mom}) reveals
a complex structure and the velocities to the south-west of the nucleus are bluer than expected from an east-west rotating disk while the velocities to the north-east are too red.
One possible explanation to this is that the nuclear inclined dusty structure drives an HCN 3--2 outflow with blueshifted emission south-west of the nucleus and redshifted emission to the north-east. This will become superposed on the underlying east-west rotation creating a convoluted velocity field. If the molecular outflow of Mrk~231 is driven by several sources (jet, disk, warped nuclear disk) the resulting outflow pattern is indeed expected to be complex.
However, there are also other potential explanations and higher resolution observations paired with modelling of the velocity structure is necessary to effectively address these issues. \\

\subsubsection{The nuclear gas mass and inclination}
\label{s:vib_kin}

The map of integrated intensities with the highest resolution (A-array map with 0.\arcsec28 $\times$  0.\arcsec25 resolution) has a peak HCN 3--2 integrated flux of 19 Jy \kms\ which translates to 4750 K \kms. 
 We could now apply an HCN to dense gas mass conversion factor to estimate the $M_{\rm dense}$ in the centre of Mrk~231. \citet{gao04} were to our knowledge the first to suggest
a global relation between the HCN 1--0 luminosity and the mass of dense gas
$M_{\rm dense}$= $10 \times L$(HCN 1--0).  Just as is the case for the CO-to-H$_2$ conversion factor this rests on the assumption on the HCN emission emerging from
a standard cloud similar to dense cores in the Milky Way.  However, \citet{garcia12} suggest that the $L$(HCN 1--0) to H$_2$ conversion factor should be lower for ULIRGs
and other HCN overluminous objects (where $L$(HCN)/$L$(CO) $\gapprox 0.1$) by a factor of 1/3.2.  
For the nucleus of Mrk 231, the HCN excitation is affected by IR and not all of its nuclear emission is emerging from the inferred warped disk (see discussion in the section above).
Instead, we can use the HCO$^+$ 3--2  emission as a proxy for an HCN emission uncontaminated by radiative excitation. The HCN conversion factor has been calibrated for the
1--0 transition so we will also assume a 3--2/1--0 lineratio of 0.5. Using the more conservative conversion factor of \citet{garcia12} (see also Sec~\ref{s:wing-mass})
we obtain a mass $M_{\rm dense}\approx 3.1 \times L$(HCN 1--0)$\approx 3.1 \times L$(HCO$^+$ 3--2) of $8 \times 10^8$ \msun\ in the inner $r$=0.\arcsec 13 and a
average $N$(H$_2$) of $1.2 \times10^{24}$ $\cmmt$.

Interestingly, this simple estimate of the nuclear gas mass is similar to the H$_2$ mass ($2.5 - 5 \times 10^8$ \msun) calculation of the core
(quiescent, QC) component of Mrk~231 by \citet{gonzalez14} from their {\it Herschel}  OH observations. We do not know the inclination of the nuclear warped disk, but if we constrain it
to have a lower limit to its dynamical mass of $5 \times 10^8$ \msun\ (the maximum H$_2$ mass from \citet{gonzalez14}) at a radius of 0.\arcsec 13 and we use the FWHM of the HCN $\nu_2$=1$f$ line to indicate that its projected rotational velocity is 130 \kms, then its inclination must be 60$^{\circ}$ or lower - which fits both the \citet{davies} and \citet{klockner03} models.

\subsubsection{Comparison to HC$_3$N and HNC}
\label{s:hnc}
The detection of vibrationally excited HCN is also interesting in the context of our recent detection of 
3mm HC$_3$N 10--9 in Mrk~231 \citep{aalto12a}. 
Models by \citet{harada13} find that HC$_3$N abundances may be enhanced substantially in warm 
($>$ 200 K) regions in the midplane of dense disks around AGNs.  \citet{lindberg} find a strong correlation between OH megamaser
activity and HC$_3$N emission implying that the regions may be co-spatial.
We therefore suggest that the HC$_3$N and HCN vibrational line emission are both centred on
the dusty central region of Mrk~231. 

We have also previously found that the velocity field of the HNC 1--0 emission in Mrk~231 is inclined with respect to the main east-west rotation.
The PA of the HNC rotation is 130$^{\circ}$ indicating that at least part of the emission seems to be associated with the proposed warped dusty 
structure \citep{aalto12a}. \citet{costagliola11} find a correlation between the HC$_3$N and HNC luminosities which is not immediately obvious from a chemical perspective.
However, both molecules can be IR pumped at mid-IR wavelengths and a possibility is that both are tracers of dust environments - just like vibrationally excited HCN.


\subsection{IR pumping of HCN?}
\label{s:pumping}

The luminous HCN emission towards many LIRGs (and in particular ULIRGs) have been used to argue that the relative dense
$n \gapprox 10^4$ $\cmmd$ gas content is higher in LIRGs than in normal galaxies. The relation between
CO/HCN 1--0 line ratio and IR luminosity has also been used as grounds to suggest that the IR luminosity is largely driven
by star formation - rather than AGN activity  \citep[e.g.][]{gao04}.  Elevated HCN/HCO$^+$ 1--0 line intensity ratios ($>1$) in the
nuclei of Seyfert galaxies \citep[e.g.][]{kohno03, usero04} and in ULIRGs \citep[e.g.][]{gracia-carpio} are suggested to be due to abundance
enhancements of HCN for example due to X-ray dominated chemistry near an AGN \citep[e.g.][]{maloney96,kohno03} or due to high temperatures
\citep[e.g.][]{harada13, kazandjian12}.

The above arguments are based on the assumption that HCN is collisionally excited.
However, it is possible that the IR pumping that leads to the detection of the $\nu_2$=1 lines also may affect the excitation of the ground state lines.
HCN has degenerate bending modes in the IR. The molecule absorbs IR-photons to the bending mode (its first vibrational state) and then it decays
back to the ground state via its $P$ or $R$ branch. In this way, a vibrational excitation may produce a change in the rotational state in the ground level and
can be treated (effectively) as a collisional excitation in the statistical equations. Thus, IR pumping excites the molecule to the higher rotational level
by a selection rule $\Delta J$=2. For a molecular medium consisting of dense clumps surrounded by lower density gas the net effect of IR pumping
may be that the HCN emission from the dense clumps gets added to by IR-pumped HCN emission from the low density gas (for a discussion of this scenario
se e.g. \citet{aalto07}) resulting in a boosted total HCN luminosity. How large this effect is depends on the balance between dense and low density gas,
the size and intensity of the IR field, and the HCN abundance.

We can make a rough estimate of the fraction of the HCN 3--2 emission in Mrk~231 that could be influenced by IR-pumping through comparing source sizes and fluxes of
the HCN and $\nu_2$=1$f$ emission. We (conservatively) assume that IR-pumping of HCN is only important where we detect the $\nu_2$=1$f$ line (inside 0.\arcsec2).
About half of the HCN 3--2 flux is found inside the A-array central beam - thus 50\% of the HCN 3--2 line emission is likely affected by IR pumping. 


\subsection{The molecular outflow}
\label{s:wing-extent-excitation}

\subsubsection{HCN excitation - dense gas in the outflow}

The HCN 3--2 line wings extend out to $\pm$750 \kms\ and has a spatial extent of at least 0.\arcsec 4 (350 pc).  The actual extent may be larger 
and the data allows for a total extent of 0.\arcsec 9.  
We can compare the flux of the 3--2 line to that of the previously published \citep{aalto12a} 1--0 line to determine the excitation
assuming that they emerge from the same region and that the line ratio is constant through the outflow. The flux in the 1--0 line wing is 0.8 (red)
and 0.6 (blue) Jy \kms. We find HCN 3--2  line wing fluxes of 2.5 (red) and 1.9 (blue) Jy km/s. The flux line ratio is then 3 for both line wings and in brightness temperature
scale $T_{\rm B}$(3--2)/$T_{\rm B}$(1--0) this corresponds to$\approx$0.35.  This value is uncertain since we have different uv-coverage and sensitivity in
the 3--2 and 1--0 data, but the ratio suggests that the emission is subthermally excited.

In the above section (Sec.~\ref{s:pumping}) we concluded that IR pumping is affecting the HCN emission in the inner 0.\arcsec2 of Mrk~231.
The line wings are more extended suggesting that they are unlikely to be significantly impacted by the nuclear IR source (apart from in the very inner part of the outflow) 
and that the HCN wing emission is therefore dominated by collisions.  The $J$=3--2 transition of HCN has a critical density of $\approx$$10^7$ $\cmmd$ and significant
HCN 3--2 emission may start to emerge at densities of $10^5$ $\cmmd$. Hence, when collisionally excited, HCN 3--2 is tracing dense gas and a 3--2/1--0 line
ratio of $\approx$0.35 suggests that the number density range for the HCN-emitting gas is $n=10^4 - 5\times 10^5$ $\cmmd$ (for kinetic temperatures $T_{\rm k}$=30 -- 100 K)
and HCN abundances $X$(HCN) of $10^9 - 10^8$ (see next section). \\

The outflow may consist of a two-phase molecular medium where the bulk of the  HCN-emitting molecular mass resides in dense  clouds and lower density, 
diffuse molecular gas is traced by low-$J$ CO.  This scenario is consistent with the apparent subthermal excitation of CO \citep{cicone12}. 

\subsubsection{Estimating the mass of dense gas in the outflow}  
\label{s:wing-mass}

The extremely bright HCN emission of the outflowing gas shows that the gas properties are very different from those where the conversion factor from CO
luminosity to H$_2$ mass was calibrated -  giant molecular clouds (GMCs) in the  disk of the Milky Way. The molecular mass is difficult to determine in an environment where the
physical conditions, structure and stability of the clouds are poorly known.   However, we can attempt to estimate the mass of dense gas through applying 
an $L$(HCN) to H$_2$ conversion factor. We adopt the same factor (determined for HCN luminous sources \citep{garcia12}) that we applied to the nuclear emission in
Section~\ref{s:vib_kin}. \citet{aalto12a} do not give a 
$L$(HCN 1--0) for the outflow, but we can estimate it here through dividing $L$(HCN 3--2) (from Tab.~\ref{t:flux}) with 0.35 (the estimated HCN 3--2/1--0 line intensity ratio
from the previous section) resulting in  $L$(HCN 1--0) $\approx 1.2\times 10^8$ K~\kms\ pc$^2$ for both the red and blue wings and the resulting $M_{\rm dense} \approx 4 \times 10^8$ \msun. (Note that the $L$(HCN 3--2) in Tab.~\ref{t:flux} is a lower limit to the real $L$(HCN 3--2) since we only include flux above a 4$\sigma$ cutoff
in the integrated intensity from which $L$(HCN 3--2) is derived.  Hence the inferred $L$(HCN 1--0) is also a lower limit.)) \\

\noindent
We can also investigate masses obtained assuming that the HCN emission is emerging from a population of dense cloud cores which we first consider to be
self-gravitating and then  we allow the line width per cloud to increase so that the clouds are no longer bound. We consider clouds of typical cloud core mass of 10~\msun\ and warm temperature of $T_{\rm k}$=50 K \citep{goldsmith87}.  \\

\noindent
{\it A. Self-gravitating clouds: } \, 
We use the RADEX non-LTE code to investigate the effect of varying HCN abundances ($X$(HCN) ranging from $10^{-8}$ to $10^{-6}$) on the brightness temperatures, luminosities
and line ratios for three model clouds of densities $n$=$3 \times 10^3$, $10^4$ and $10^5$ $\cmmd$ (where $n$=$3 \times 10^3$ is the lowest density for which a solution could be found for the given range of  $X$(HCN)) .  In Tab.~\ref{t:clump} we present a few examples of model clouds that fulfil the observed line ratios (within the errors) keeping in mind that they are illustrations of possible solutions. To get the total mass of dense gas from the ensemble of model clouds we divided the observed $L$(HCN 3--2) by the luminosity per model cloud. For high abundances ($X$(HCN)=$10^{-6}$) we can find solutions to fit the observed line ratios for the lower density model clouds and total dense gas masses are $M_{\rm dense}$=$4.4 \times 10^8$ and  $3.5 \times 10^8$ \msun. 
When we lower the HCN abundances to more normal values of $10^{-8}$ we find solutions only for the high density model cloud. The luminosity per cloud is lower due
to the smaller radius and $M_{\rm dense}$=$1.7 \times 10^9$ \msun. 

We can conclude that an ensemble of self-gravitating clouds with a high HCN abundance reproduces
the mass of dense gas we obtain quite well when adopting the $L$(HCN) to $M_{\rm dense}$ conversion factor at the beginning of the section.  This is not surprising since
inherent in the conversion factor lies an assumption of self-gravitating clouds.  In addition the modified $L$(HCN) conversion factor for luminous HCN sources
(see Sect.~\ref{s:vib_kin}) attempts to scale down the masses obtained for extremely luminous (relative to CO) HCN sources. In our RADEX model we achieve this through adopting a large HCN abundance for self-gravitating clouds. For lower HCN abundances we require denser clouds to fit the line ratios and the resulting total mass goes up since the
luminosity per cloud goes down.  It is also generally true that if we increase the kinetic temperature the luminosity per cloud increases reducing the total molecular mass. \\

\begin{table}
\caption{\label{t:clump} Model HCN  3--2 luminosities - Self-gravitating cloud$^a$}
\begin{tabular}{lcccccc} 
\hline
\hline\\[0.001 pt]
 & $n$ & $X$ & R & $\Delta V$ &  $N$(HCN) & $L$\\
\hline \\[0.001ex]
& [ cm$^{-3}$ ] &  &  [pc] & [\kms] &  [cm$^{-2}$] & [K~\kms\ pc$^2$]\\
\hline \\ 
1. & $3\times 10^3$ & $10^{-6}$ & 0.25 & 0.4 &  $5 \times 10^{15}$ & 0.95 \\
2. & $10^4$ & $10^{-6}$ & 0.17 & 0.5 &  $10^{16}$ & 1.2 \\
3. & $10^5$ & $10^{-8}$ & 0.08 & 0.7 &  $5 \times 10^{14}$& 0.25 \\
\\
\hline \\
\end{tabular} 

a) We are considering simple self-gravitating clouds where $M_{\rm vir}$=($R (\Delta V)^2/G$ (virial mass $M_{\rm vir}$, radius $R$ and line width $\Delta V$). 
We assume $M_{\rm vir}$=10\msun\ and temperature $T_{\rm k}$=50 K for every cloud. For model 1. $N$(H$_2$)=$2.5 \times 10^{21}$ $\cmmt$, for 2. $N$(H$_2$)=$10^{22}$ $\cmmt$, and for 3. $N$(H$_2$)=$5 \times 10^{22}$ $\cmmt$. $X$ is the HCN abundance relative to H$_2$, $N$(HCN) the HCN column density and $L$ is the HCN 3--2 luminosity: $\pi R^2 I$(HCN)$\approx$$\pi R^2T_{\rm B}({\rm HCN})1.06\Delta V$. The model clouds fulfil: HCN 3--2/1--0 = 0.3 to 0.4 and CO/HCN 1--0 = 1 to 3. The three HCN 1--0 hyperfine structures are summed in the intensity of the HCN 1-0 line since the inferred line width per model cloud is smaller than the velocity separation between the hyperfine structures. (We do not account for the radiative transfer effects in the ensemble of clouds which would lower the resulting HCN 1--0 intensity in relation to HCN 3--2 and CO 1--0). CO/HCN 1--0 ratios of unity are allowed to account for the possibility that not all of the CO flux is emerging from these HCN-emitting clouds.

\end{table}


\noindent
{\it B. Non self-gravitating clouds: } \, We can simulate an ensemble of unbound clouds through increasing the line width $\Delta V$ per cloud. Now the clouds will be ``overluminous"  with respect to their mass compared to self-gravitating clouds. We can test this scenario through taking the first model cloud (1) in Tab.~\ref{t:clump} and increase the linewidth by a factor of twenty - keeping all other parameters intact. The increase in $\Delta V$ results in a drop in HCN 3--2 brightness temperature by a factor of 13, so the net increase in luminosity per cloud is 1.5. Now, however, the HCN 3--2/1--0 line ratio becomes too low and the HCN 1--0 emission too faint compared to CO 1--0 (see \citep{aalto12a}). We can compensate for this through increasing the HCN abundance further to $>10^{-5}$ (which is a very high abundance). As a consequence, low density ($n \lapprox 10^4$ $\cmmd$) solutions need extreme HCN abundances to compensate for the loss in optical depth and excitation caused by the increased line width.
It is easier to maintain a high HCN brightness in denser clouds since the excitation is less subthermal. For example, increasing $\Delta V$ by a factor of 10  and $X$(HCN) to  $10^{-6}$ for model cloud  3. in Tab.~\ref{t:clump} will result in a factor of 23 more HCN 3--2 luminosity per cloud reducing $M_{\rm dense}$
to $\approx 7 \times 10^7$.   

\noindent
To summarize, for a medium consisting of self-gravitating dense clouds we find a mass of dense gas $M_{\rm dense} \approx 4 \times 10^8$ \msun\ - which will accomodate most reasonable RADEX solutions as well as the mass when applying the HCN to H$_2$ conversion factor at the beginning of this section. An ensemble consisting of non-self-gravitating clouds may have lower molecular masses by up to an order of magnitude for the same HCN luminosity, but it requires $X$(HCN)$\gapprox 10^{-6}$ to maintain the large HCN brightness in relation to CO. 
Abundances of $X$(HCN)$\approx 10^{-8}$ would be considered rather typical for Galactic GMCs and larger scale star forming regions \citep{irvine87}. Enhanced abundances
of HCN are found in warm, dense  regions, for example in hot cores or shocks \citet{jorgensen04,tafalla10} where HCN may be enhanced by factors 10 - 100 over HCO$^+$. 
\citet{harada13} also suggest that HCN is enhanced in X-ray dominated/illuminated regions (XDRs) in AGNs where the HCN-enhancement is caused by the high temperatures in the
XDRs. Possible HCN abundance enhancements in the outflow of Mrk~231 would likely stem from shocks potentially originating from the radio jet interacting with the outflowing gas and/or multiple phases of the outflow colliding. 

\noindent
Finally, the distribution of the dense clouds is unlikely to be smooth within the outflow and even viewing them as individual clouds could be wrong since they may (for example)
consitute dense condensations along filaments or along walls of bipolar structures. As is already evident in the pV diagrams (see Fig.~\ref{f:wings}) the wing emission is
clumpy in velocity space showing that the gas is not smoothly distributed. We will discuss this further in a subsequent paper (Lindberg et. al in prep.) We adopt the mass of self-gravitating clouds (and the modified conversion factor) $M_{\rm dense} \approx 4 \times 10^8$ \msun\ as an upper limit to the dense molecular gas mass in the outflow.

\subsubsection{Can dense clouds survive in the outflow?}  

Results from recent simulations by \citet{gabor14} of the physical conditions of thermal AGN outflows imply that the outflow consists mostly of hot ($10^7$ K) gas
of low density. Dense clumps may be entrained by these hot flows but they are warmer and less dense than the disk gas. \citet{gabor14}  also find that only a small mass
fraction of the outflow consists of molecular gas and that the AGN feedback does not strongly affect the surrounding disk since it only impacts the diffuse gas.
Models by \citet{zubovas14} instead imply that the outflows can not remain in a hot-gas phase, but forms a two-phase medium where cold clumps are mixed in with 
the tenuous gas and most of the mass is in the dense clumps. The authors suggest that cooling should lead to star formation in
the outflow where the star formation efficiency (SFE)  is high and quickly removes dense gas from the outflow. We also note that there is observational evidence that
molecules may reform in fast outflows as seen in the supernova remnant Cas~A where  dense CO clumps are detected in the region of the reverse shock \citep{wallstrom13}.

We note that the low-$J$ CO emission seems to extend out to twice the radial distance than the
HCN outflow \citep{feruglio10,cicone12} which suggests that the dense gas is either more difficult to drive to large distances. Alternatively the dense gas evaporates or is turned into stars.
Our results therefore seem to (tentatively) support the models of \citet{zubovas14} over those of \citet{gabor14}  - or that the outflow of Mrk~231 is atypical for AGN outflows.  
The latter seems unlikely since luminous \cite{wallstrom13}HCN 4--3 emission is detected in the AGN-driven outflow of NGC1068 \citep{garcia14} together with bright CO 6--5 emission
indicating the presence of dense clouds. The outflow is likely powered by the AGN jets and also for Mrk231 it is  possible
that the dense gas is not primarily being driven out by a thermal wind from the AGN, but by the radio jet or by radiation pressure from the hot dusty warped disk.

\subsubsection{What is powering the dense outflow?}
\label{s:dusty-flow}

The mechanical luminosity of the outflowing dense gas is challenging to determine since we have significant error bars on its mass. We can take the upper limit
of $4 \times 10^8$ \msun\ (Sec.~\ref{s:wing-mass}) and let it move at a (constant) velocity of 750 \kms.  Let the total kinetic energy be
$E_{\rm kin}=E_{\rm bulk} + E_{\rm turbulent}$ \citep{veilleux01} where $E_{\rm bulk}=\Sigma (1/2) \delta m_i (v_i)^2=(1/2)M_{\rm out} \times V_{\rm out}^2$
is then $\approx 2.4 \times 10^{57}$ erg. We cannot measure a reliable limit of the turbulence of the gas, but assume that the contribution is significantly smaller
than that of the bulk movement. We do not know the radial extent of the outflow, but a lower limit is 350 pc which results in an upper limit to the mechanical luminosity
of $L_{\rm mech}=1.5\times10^{44}$ ergs s$^{-1}$ or $4 \times 10^{10}$ \lsun.  The outflow rate $\dot{M}_{\rm out}$=800 \msun\ yr$^{-1}$ and the momentum
flux $\dot{P}=\dot{M} \times v =4 \times 10^{36}$ g cm s$^{-2}$. This is about  12$L_{\rm AGN}/c$ (for $L_{\rm AGN}=2.8 \times 10^{12}$ \lsun).   However,
if the dense clouds in the outflow are not self-gravitating (or near self-gravitating) the mass of dense gas in the outflow may be one order of magnitude lower with a corresponding 
reduction in kinetic energy and momentum flux. Furthermore, if the outflow is decelerating so that a portion of the gas has lower velocities then estimates will
come down further.

From the dust models of \citet{gonzalez10} the luminosity of the hot (150-400 K) compact ($r$=23 pc) dust component) and the warm (95 K) component ($r$=120 pc) is estimated to be
$L_{\rm dust}=L_{\rm hot} + L_{\rm warm}=7.5 \times 10^{11} + 1.9 \times 10^{12}=2.65 \times 10^{12}$ \lsun.
Simulations of clumpy outflows by \citet{roth} suggest that momentum driven outflows may exceed $L/c$, but only by factors of up to 5 (although this seems to be dependent on the 
scale height of their model disk). Thus for the maximum mass estimate of
the dense outflow  radiation pressure will not play a major role in driving out the gas - instead winds from the AGN and/or boosting from the jets are the main candidates for the
powering of the outflow.  However, if the $\dot{P}$ is instead  1.2$L_{\rm AGN}/c$ (for the lower mass (and/or lower velocity) estimate) then radiation pressure may well be the main power source.


\subsubsection{Comparing to the OH outflow}
\label{s:OH-flow}

{\it Herschel} observations of IR OH towards Mrk~231 reveal two outflowing components: one low excitation feature with velocities 200 - 800 \kms\ (LEC component) and one high
excitation feature (HVC component) with velocities up to $\approx$1500 \kms\ \citep{gonzalez14}. The dense HCN 3--2 outflow studied here has velocities similar to the OH LEC component.
Our estimated mechanical luminosities agree within a factor of 2-3. The gas densities required to excite the HCN 3--2 line in the outflow, however, are much higher than the averaged
values derived from OH and CO, strongly pointing towards a clumpy structure in the HCN outflow.

We see no sign of the high $\approx$1500 \kms\ velocities seen in the highly excited OH absorption. This component seems to be nuclear and radiatively excited and \citet{gonzalez14}
suggest that it may probe an interclump medium of the torus itself that is flowing past the dense clumps and hence would move faster than the dense gas. It is suggested that
interaction with the dense clumps would decelerate the outflowing gas with increasing radius. The OH HVC component is radiatively excited but would not be seen as a radiatively
excited HCN component in emission due to too low column density - and too small spatial extent. While absorption measurements of high lying lines towards very warm continuum sources are primarily sensitive to the rates ($\dot{M}$, $\dot{P}$, $L_{\rm mech}$), emission measurements are more sensitive to integral values ($M_{\rm gas}$, $E_{\rm kin}$) \citep{gonzalez14}.

\subsection{Mrk~231 as a dust-obscured QSO}
\label{s:dusty-qso}

The detection of the HCN $J$=3--2 $\nu_2$=1, $l$=1$f$ vibrational line shows that the HCN molecule is abundant even in extreme, hot nuclear environments. Since
the HCN vibrational line is excited by intense IR dust emission it also suggests that the QSO has not yet shed itself of its nuclear shroud
and this dusty, warm region may represent the final stage of the obscured,  X-ray absorbed \citep{page} accretion phase of the QSO. 
The lack of mid-IR AGN signatures may indicate that the coronal region around the AGN is blocked from our line of sight \citep{armus07}. Interestingly, the
far-ultraviolet spectrum shows a lack of absorption signatures \citep{veilleux13} and recent X-ray studies by \citet{teng14} suggest that the AGN is being viewed through a patchy
and Compton-thin ($N$(H$_2$) $\approx 10^{23}$ $\cmmt$) column. The authors propose that the AGN is intrinsically X-ray faint thus explaining the lack of mid-IR signatures.

The $\nu_2$=1$f$ HCN line is excited by mid-IR dust emission, and its optical depth will peak at
the location of the dust source.  The position of the QSO core agrees (within the errors) with the peak of the host dust, which is consistent with the notion that there is an
inner radius of the torus/disk is close (20~pc) to the AGN \citep{klockner03,richards05}. Our estimated $N$(H$_2$) in the inner diameter of 0.\arcsec 13 of $1.2 \times10^{24}$ $\cmmt$ is an average value that is consistent with the scenario where the QSO has blown a hole in the central gas and dust distribution, creating a torus-like structure inside the warped nuclear disk. 

Further observations may reveal to what degree the warped  dusty disk is collimating the outflowing dense gas. The inclination of the nuclear structure is
not constrained and the structure of the dense outflow seems complex, meaning that even higher spatial resolution is needed to disentangle the nuclear and outflow dynamics. 
In addition, more studies will also reveal whether there is a massive inflow of gas to replenish the central region feeding the powerful molecular outflow.   If so, Mrk~231 may
continue to be a dust-obscured QSO until all gas reservoirs have been exhausted, or until the inflow process is halted. If there is no significant inflow and if the molecular
outflow indeed originates in the inner region near the AGN, the available bulk of gas of $\approx 5\times 10^8$ \msun\ will be emptied within the extremely short time
of 0.5 Myr, assuming a steady flow and adopting the upper limit on the mass estimate from Sec.~\ref{s:wing-mass}.


\section{Conclusions}



We detected luminous emission from HCN and HCO$^+$ $J$=3--2 in the main disk of Mrk~231. The HCN 3--2 emission is concentrated towards the inner region 
(FWHM $r$=0.\arcsec 17 (150 pc)) but disk emission for both HCN and HCO$^+$ extends out to at least 600 pc - with southern and western emission features of 1~kpc.  Line wings are detected for HCN where velocities are found to be similar to those found for CO and HCN $J$=1--0 ($ \pm 750$ \kms). No line wings are found for HCO$^+$ 3--2.
We find that both the blue- and redshifted HCN $J$=3--2 line wings are spatially extended  $>$0.\arcsec 4 ($\gapprox$350 pc) centred on the nucleus. 256~GHz continuum is detected at 46 mJy - more than half of it is extended on the scales of the main disk. Emission from the reactive ion HOC$^+$ is tentatively detected  with a HCO$^+$/HOC$^+$ $J$=3--2  line intensity ratio of 10-20.

We detected the HCN $J$=3--2 $\nu_2$=1$f$ vibrational line for the first time in Mrk~231. The HCN $\nu_2$=1$f$ line emission is compact ($r \lapprox$ 0.\arcsec 1 (90 pc)) and centred on the nucleus where it is excited by bright mid-IR 14 $\mu$m continuum to a vibrational temperature $T_{\rm vib}$=200-400 K (assuming co-spatial HCN $\nu$=0 and $\nu_2$=1$f$ ) . The HCN $\nu_2$=1$f$ velocity field line of nodes are oriented with a PA of 155$^{\circ} \pm 10^{\circ}$ - strongly inclined to the rotation of the main disk.  We suggest that the HCN $J$=3--2 $\nu_2$=1$f$ line is emerging from the inner dusty region of a warped disk, as seen also in stellar absorption lines and in OH megamaser emission.  The nuclear ($r \lapprox$ 0.\arcsec 1) molecular mass is estimated to $8 \times10^8$ \msun\ and the average column density $N$(H$_2$)=$1.2 \times10^{24}$ $\cmmt$. The detection of the HCN $J$=3--2 $\nu_2$=1$f$ vibrational line is consistent with the notion of Mrk~231 as a QSO in the final stages of its dust-obscured phase.

The HCN $J$=3--2 $\nu_2$=1$f$ vibrational line also reveals that the HCN excitation is affected by IR radiative excitation and that IR pumping may also affect the rotational levels of
the vibrational ground state, hence the excitation of the whole molecule. A vibrational excitation may produce a change in the rotational state in the ground level and
can be treated (effectively) as a collisional excitation in the statistical equations. We estimate that 50\% of the HCN 3--2 emission in the main disk may be influenced by radiative
excitation.

The HCN emission from the line wings is extended, so we concluded that IR pumping does not strongly influence the excitation (apart from possibly in the inner part). Instead, we propose that most of the HCN emission in the outflow is collisionally excited. Line ratios indicate that the emission is emerging in dense gas $n=10^4 - 5 \times 10^5$ $\cmmd$, 
but elevated HCN abundances may also allow for emission originating from more moderate densities. The HCN abundance in the outflow seems high and $X$(HCN) may be $10^{-6}$. Elevated HCN abundances are found in warm, dense gas in star forming and shocked regions in the Galaxy. Models also indicate that HCN may be very abundant in X-ray irradiated warm gas in AGN disks. We estimated an upper limit to the dense mass of $4 \times 10^8$ \msun\ with an outflow rate $\dot{M}_{\rm out}$=800 \msun\ yr$^{-1}$ and a momentum flux $\dot{P} \lapprox$ 12$L_{\rm AGN}$. If the dense gas clouds are not self-gravitating, the mass and momentum flux will be lower.


\begin{acknowledgements}
 We thank the IRAM PdBI staff for excellent support. SA thanks the Swedish National Science Council for grant (Dnr. 621-2011-4143) support.
\end{acknowledgements}

\bibliographystyle{aa}
\bibliography{mrk231_ref}

\begin{thebibliography}{53}
\expandafter\ifx\csname natexlab\endcsname\relax\def\natexlab#1{#1}\fi

\bibitem[{{Aalto} {et~al.}(2012){Aalto}, {Garcia-Burillo}, {Muller}, {Winters},
  {van der Werf}, {Henkel}, {Costagliola}, \& {Neri}}]{aalto12a}
{Aalto}, S., {Garcia-Burillo}, S., {Muller}, S., {et~al.} 2012, \aap, 537, A44

\bibitem[{{Aalto} {et~al.}(2007){Aalto}, {Spaans}, {Wiedner}, \&
  {H{\"u}ttemeister}}]{aalto07}
{Aalto}, S., {Spaans}, M., {Wiedner}, M.~C., \& {H{\"u}ttemeister}, S. 2007,
  \aap, 464, 193

\bibitem[{{Armus} {et~al.}(2007){Armus}, {Charmandaris}, {Bernard-Salas},
  {Spoon}, {Marshall}, {Higdon}, {Desai}, {Teplitz}, {Hao}, {Devost}, {Brandl},
  {Wu}, {Sloan}, {Soifer}, {Houck}, \& {Herter}}]{armus07}
{Armus}, L., {Charmandaris}, V., {Bernard-Salas}, J., {et~al.} 2007, \apj, 656,
  148

\bibitem[{{Bryant} \& {Scoville}(1996)}]{bryant}
{Bryant}, P.~M. \& {Scoville}, N.~Z. 1996, \apj, 457, 678

\bibitem[{{Carilli} {et~al.}(1998){Carilli}, {Wrobel}, \&
  {Ulvestad}}]{carilli98}
{Carilli}, C.~L., {Wrobel}, J.~M., \& {Ulvestad}, J.~S. 1998, \aj, 115, 928

\bibitem[{{Cicone} {et~al.}(2012){Cicone}, {Feruglio}, {Maiolino}, {Fiore},
  {Piconcelli}, {Menci}, {Aussel}, \& {Sturm}}]{cicone12}
{Cicone}, C., {Feruglio}, C., {Maiolino}, R., {et~al.} 2012, \aap, 543, A99

\bibitem[{{Costagliola} {et~al.}(2011){Costagliola}, {Aalto}, {Rodriguez},
  {Muller}, {Spoon}, {Mart{\'{\i}}n}, {Per{\'e}z-Torres}, {Alberdi},
  {Lindberg}, {Batejat}, {J{\"u}tte}, {van der Werf}, \&
  {Lahuis}}]{costagliola11}
{Costagliola}, F., {Aalto}, S., {Rodriguez}, M.~I., {et~al.} 2011, \aap, 528,
  A30+

\bibitem[{{Davies} {et~al.}(2004){Davies}, {Tacconi}, \& {Genzel}}]{davies}
{Davies}, R.~I., {Tacconi}, L.~J., \& {Genzel}, R. 2004, \apj, 613, 781

\bibitem[{{Downes} \& {Solomon}(1998)}]{downes98}
{Downes}, D. \& {Solomon}, P.~M. 1998, \apj, 507, 615

\bibitem[{{Feruglio} {et~al.}(2010){Feruglio}, {Maiolino}, {Piconcelli},
  {Menci}, {Aussel}, {Lamastra}, \& {Fiore}}]{feruglio10}
{Feruglio}, C., {Maiolino}, R., {Piconcelli}, E., {et~al.} 2010, \aap, 518,
  L155+

\bibitem[{{Fischer} {et~al.}(2010){Fischer}, {Sturm}, {Gonz{\'a}lez-Alfonso},
  {Graci{\'a}-Carpio}, {Hailey-Dunsheath}, {Poglitsch}, {Contursi}, {Lutz},
  {Genzel}, {Sternberg}, {Verma}, \& {Tacconi}}]{fischer10}
{Fischer}, J., {Sturm}, E., {Gonz{\'a}lez-Alfonso}, E., {et~al.} 2010, \aap,
  518, L41

\bibitem[{{Gabor} \& {Bournaud}(2014)}]{gabor14}
{Gabor}, J.~M. \& {Bournaud}, F. 2014, ArXiv e-prints

\bibitem[{{Gallagher} {et~al.}(2002){Gallagher}, {Brandt}, {Chartas},
  {Garmire}, \& {Sambruna}}]{gallagher02}
{Gallagher}, S.~C., {Brandt}, W.~N., {Chartas}, G., {Garmire}, G.~P., \&
  {Sambruna}, R.~M. 2002, \apj, 569, 655

\bibitem[{{Gao} \& {Solomon}(2004)}]{gao04}
{Gao}, Y. \& {Solomon}, P.~M. 2004, \apj, 606, 271

\bibitem[{{Garc{\'{\i}}a-Burillo} {et~al.}(2014){Garc{\'{\i}}a-Burillo},
  {Combes}, {Usero}, {Aalto}, {Krips}, {Viti}, {Alonso-Herrero}, {Hunt},
  {Schinnerer}, {Baker}, {Boone}, {Casasola}, {Colina}, {Costagliola},
  {Eckart}, {Fuente}, {Henkel}, {Labiano}, {Mart{\'{\i}}n}, {M{\'a}rquez},
  {Muller}, {Planesas}, {Ramos Almeida}, {Spaans}, {Tacconi}, \& {van der
  Werf}}]{garcia14}
{Garc{\'{\i}}a-Burillo}, S., {Combes}, F., {Usero}, A., {et~al.} 2014, \aap,
  567, A125

\bibitem[{{Garc{\'{\i}}a-Burillo} {et~al.}(2012){Garc{\'{\i}}a-Burillo},
  {Usero}, {Alonso-Herrero}, {Graci{\'a}-Carpio}, {Pereira-Santaella},
  {Colina}, {Planesas}, \& {Arribas}}]{garcia12}
{Garc{\'{\i}}a-Burillo}, S., {Usero}, A., {Alonso-Herrero}, A., {et~al.} 2012,
  \aap, 539, A8

\bibitem[{{Goldsmith}(1987)}]{goldsmith87}
{Goldsmith}, P.~F. 1987, in Astrophysics and Space Science Library, Vol. 134,
  Interstellar Processes, ed. D.~J. {Hollenbach} \& H.~A. {Thronson}, Jr.,
  51--70

\bibitem[{{Gonz{\'a}lez-Alfonso} {et~al.}(2014){Gonz{\'a}lez-Alfonso},
  {Fischer}, {Graci{\'a}-Carpio}, {Falstad}, {Sturm}, {Mel{\'e}ndez}, {Spoon},
  {Verma}, {Davies}, {Lutz}, {Aalto}, {Polisensky}, {Poglitsch}, {Veilleux}, \&
  {Contursi}}]{gonzalez14}
{Gonz{\'a}lez-Alfonso}, E., {Fischer}, J., {Graci{\'a}-Carpio}, J., {et~al.}
  2014, \aap, 561, A27

\bibitem[{{Gonz{\'a}lez-Alfonso} {et~al.}(2010){Gonz{\'a}lez-Alfonso},
  {Fischer}, {Isaak}, {Rykala}, {Savini}, {Spaans}, {van der Werf},
  {Meijerink}, {Israel}, {Loenen}, {Vlahakis}, {Smith}, {Charmandaris},
  {Aalto}, {Henkel}, {Wei{\ss}}, {Walter}, {Greve}, {Mart{\'{\i}}n-Pintado},
  {Naylor}, {Spinoglio}, {Veilleux}, {Harris}, {Armus}, {Lord}, {Mazzarella},
  {Xilouris}, {Sanders}, {Dasyra}, {Wiedner}, {Kramer}, {Papadopoulos},
  {Stacey}, {Evans}, \& {Gao}}]{gonzalez10}
{Gonz{\'a}lez-Alfonso}, E., {Fischer}, J., {Isaak}, K., {et~al.} 2010, \aap,
  518, L43+

\bibitem[{{Graci{\'a}-Carpio} {et~al.}(2006){Graci{\'a}-Carpio},
  {Garc{\'{\i}}a-Burillo}, {Planesas}, \& {Colina}}]{gracia-carpio}
{Graci{\'a}-Carpio}, J., {Garc{\'{\i}}a-Burillo}, S., {Planesas}, P., \&
  {Colina}, L. 2006, \apjl, 640, L135

\bibitem[{{Harada} {et~al.}(2013){Harada}, {Thompson}, \& {Herbst}}]{harada13}
{Harada}, N., {Thompson}, T.~A., \& {Herbst}, E. 2013, \apj, 765, 108

\bibitem[{{Imanishi} \& {Nakanishi}(2013)}]{imanishi13}
{Imanishi}, M. \& {Nakanishi}, K. 2013, \aj, 146, 91

\bibitem[{{Irvine} {et~al.}(1987){Irvine}, {Goldsmith}, \&
  {Hjalmarson}}]{irvine87}
{Irvine}, W.~M., {Goldsmith}, P.~F., \& {Hjalmarson}, A. 1987, in Astrophysics
  and Space Science Library, Vol. 134, Interstellar Processes, ed. D.~J.
  {Hollenbach} \& H.~A. {Thronson}, Jr., 561--609

\bibitem[{{J{\o}rgensen} {et~al.}(2004){J{\o}rgensen}, {Hogerheijde}, {Blake},
  {van Dishoeck}, {Mundy}, \& {Sch{\"o}ier}}]{jorgensen04}
{J{\o}rgensen}, J.~K., {Hogerheijde}, M.~R., {Blake}, G.~A., {et~al.} 2004,
  \aap, 415, 1021

\bibitem[{{Kazandjian} {et~al.}(2012){Kazandjian}, {Meijerink}, {Pelupessy},
  {Israel}, \& {Spaans}}]{kazandjian12}
{Kazandjian}, M.~V., {Meijerink}, R., {Pelupessy}, I., {Israel}, F.~P., \&
  {Spaans}, M. 2012, \aap, 542, A65

\bibitem[{{Kl{\"o}ckner} {et~al.}(2003){Kl{\"o}ckner}, {Baan}, \&
  {Garrett}}]{klockner03}
{Kl{\"o}ckner}, H.-R., {Baan}, W.~A., \& {Garrett}, M.~A. 2003, \nat, 421, 821

\bibitem[{{Kohno}(2003)}]{kohno03}
{Kohno}, K. 2003, in Astronomical Society of the Pacific Conference Series,
  Vol. 289, The Proceedings of the IAU 8th Asian-Pacific Regional Meeting,
  Volume 1, ed. S.~{Ikeuchi}, J.~{Hearnshaw}, \& T.~{Hanawa}, 349--352

\bibitem[{{Lahuis} {et~al.}(2007){Lahuis}, {Spoon}, {Tielens}, {Doty}, {Armus},
  {Charmandaris}, {Houck}, {St{\"a}uber}, \& {van Dishoeck}}]{lahuis07}
{Lahuis}, F., {Spoon}, H.~W.~W., {Tielens}, A.~G.~G.~M., {et~al.} 2007, \apj,
  659, 296

\bibitem[{{Lindberg} {et~al.}(2011){Lindberg}, {Aalto}, {Costagliola},
  {P{\'e}rez-Beaupuits}, {Monje}, \& {Muller}}]{lindberg}
{Lindberg}, J.~E., {Aalto}, S., {Costagliola}, F., {et~al.} 2011, \aap, 527,
  A150

\bibitem[{{Lipari} {et~al.}(2009){Lipari}, {Sanchez}, {Bergmann}, {Terlevich},
  {Garcia-Lorenzo}, {Punsly}, {Mediavilla}, {Taniguchi}, {Ajiki}, {Zheng},
  {Acosta}, \& {Jahnke}}]{lipari09}
{Lipari}, S., {Sanchez}, S.~F., {Bergmann}, M., {et~al.} 2009, \mnras, 392,
  1295

\bibitem[{{L{\'{\i}}pari} {et~al.}(2005){L{\'{\i}}pari}, {Terlevich}, {Zheng},
  {Garcia-Lorenzo}, {Sanchez}, \& {Bergmann}}]{lipari05}
{L{\'{\i}}pari}, S., {Terlevich}, R., {Zheng}, W., {et~al.} 2005, \mnras, 360,
  416

\bibitem[{{Lonsdale} {et~al.}(2003){Lonsdale}, {Lonsdale}, {Smith}, \&
  {Diamond}}]{lonsdale03}
{Lonsdale}, C.~J., {Lonsdale}, C.~J., {Smith}, H.~E., \& {Diamond}, P.~J. 2003,
  \apj, 592, 804

\bibitem[{{Maloney} {et~al.}(1996){Maloney}, {Hollenbach}, \&
  {Tielens}}]{maloney96}
{Maloney}, P.~R., {Hollenbach}, D.~J., \& {Tielens}, A.~G.~G.~M. 1996, \apj,
  466, 561

\bibitem[{{Mills} {et~al.}(2013){Mills}, {G{\"u}sten}, {Requena-Torres}, \&
  {Morris}}]{mills13}
{Mills}, E.~A.~C., {G{\"u}sten}, R., {Requena-Torres}, M.~A., \& {Morris},
  M.~R. 2013, \apj, 779, 47

\bibitem[{{Page} {et~al.}(2004){Page}, {Stevens}, {Ivison}, \&
  {Carrera}}]{page}
{Page}, M.~J., {Stevens}, J.~A., {Ivison}, R.~J., \& {Carrera}, F.~J. 2004,
  \apjl, 611, L85

\bibitem[{{Patnaik} {et~al.}(1992){Patnaik}, {Browne}, {Wilkinson}, \&
  {Wrobel}}]{patnaik92}
{Patnaik}, A.~R., {Browne}, I.~W.~A., {Wilkinson}, P.~N., \& {Wrobel}, J.~M.
  1992, \mnras, 254, 655

\bibitem[{{Richards} {et~al.}(2005){Richards}, {Knapen}, {Yates}, {Cohen},
  {Collett}, {Wright}, {Gray}, \& {Field}}]{richards05}
{Richards}, A.~M.~S., {Knapen}, J.~H., {Yates}, J.~A., {et~al.} 2005, \mnras,
  364, 353

\bibitem[{{Roth} {et~al.}(2012){Roth}, {Kasen}, {Hopkins}, \&
  {Quataert}}]{roth}
{Roth}, N., {Kasen}, D., {Hopkins}, P.~F., \& {Quataert}, E. 2012, \apj, 759,
  36

\bibitem[{{Rupke} \& {Veilleux}(2011)}]{rupke11}
{Rupke}, D.~S.~N. \& {Veilleux}, S. 2011, \apjl, 729, L27+

\bibitem[{{Sakamoto} {et~al.}(2010){Sakamoto}, {Aalto}, {Evans}, {Wiedner}, \&
  {Wilner}}]{sakamoto10}
{Sakamoto}, K., {Aalto}, S., {Evans}, A.~S., {Wiedner}, M.~C., \& {Wilner},
  D.~J. 2010, \apjl, 725, L228

\bibitem[{{Soifer} {et~al.}(2000){Soifer}, {Neugebauer}, {Matthews}, {Egami},
  {Becklin}, {Weinberger}, {Ressler}, {Werner}, {Evans}, {Scoville}, {Surace},
  \& {Condon}}]{soifer}
{Soifer}, B.~T., {Neugebauer}, G., {Matthews}, K., {et~al.} 2000, \aj, 119, 509

\bibitem[{{Solomon} {et~al.}(1992){Solomon}, {Downes}, \&
  {Radford}}]{solomon92}
{Solomon}, P.~M., {Downes}, D., \& {Radford}, S.~J.~E. 1992, \apjl, 387, L55

\bibitem[{{Tafalla} {et~al.}(2010){Tafalla}, {Santiago-Garc{\'{\i}}a}, {Hacar},
  \& {Bachiller}}]{tafalla10}
{Tafalla}, M., {Santiago-Garc{\'{\i}}a}, J., {Hacar}, A., \& {Bachiller}, R.
  2010, \aap, 522, A91+

\bibitem[{{Taylor} {et~al.}(1999){Taylor}, {Silver}, {Ulvestad}, \&
  {Carilli}}]{taylor99}
{Taylor}, G.~B., {Silver}, C.~S., {Ulvestad}, J.~S., \& {Carilli}, C.~L. 1999,
  \apj, 519, 185

\bibitem[{{Teng} {et~al.}(2014){Teng}, {Brandt}, {Harrison}, {Luo},
  {Alexander}, {Bauer}, {Boggs}, {Christensen}, {Comastri}, {Craig}, {Fabian},
  {Farrah}, {Fiore}, {Gandhi}, {Grefenstette}, {Hailey}, {Hickox}, {Madsen},
  {Ptak}, {Rigby}, {Risaliti}, {Saez}, {Stern}, {Veilleux}, {Walton}, {Wik}, \&
  {Zhang}}]{teng14}
{Teng}, S.~H., {Brandt}, W.~N., {Harrison}, F.~A., {et~al.} 2014, \apj, 785, 19

\bibitem[{{Ulvestad} {et~al.}(1999){Ulvestad}, {Wrobel}, {Roy}, {Wilson},
  {Falcke}, \& {Krichbaum}}]{ulvestad99}
{Ulvestad}, J.~S., {Wrobel}, J.~M., {Roy}, A.~L., {et~al.} 1999, \apjl, 517,
  L81

\bibitem[{{Usero} {et~al.}(2004){Usero}, {Garc{\'{\i}}a-Burillo}, {Fuente},
  {Mart{\'{\i}}n-Pintado}, \& {Rodr{\'{\i}}guez-Fern{\'a}ndez}}]{usero04}
{Usero}, A., {Garc{\'{\i}}a-Burillo}, S., {Fuente}, A.,
  {Mart{\'{\i}}n-Pintado}, J., \& {Rodr{\'{\i}}guez-Fern{\'a}ndez}, N.~J. 2004,
  \aap, 419, 897

\bibitem[{{van der Tak} {et~al.}(2007){van der Tak}, {Black}, {Sch{\"o}ier},
  {Jansen}, \& {van Dishoeck}}]{vandertak}
{van der Tak}, F.~F.~S., {Black}, J.~H., {Sch{\"o}ier}, F.~L., {Jansen}, D.~J.,
  \& {van Dishoeck}, E.~F. 2007, \aap, 468, 627

\bibitem[{{Veilleux} {et~al.}(2001){Veilleux}, {Shopbell}, \&
  {Miller}}]{veilleux01}
{Veilleux}, S., {Shopbell}, P.~L., \& {Miller}, S.~T. 2001, \aj, 121, 198

\bibitem[{{Veilleux} {et~al.}(2013){Veilleux}, {Trippe}, {Hamann}, {Rupke},
  {Tripp}, {Netzer}, {Lutz}, {Sembach}, {Krug}, {Teng}, {Genzel}, {Maiolino},
  {Sturm}, \& {Tacconi}}]{veilleux13}
{Veilleux}, S., {Trippe}, M., {Hamann}, F., {et~al.} 2013, \apj, 764, 15

\bibitem[{{Wallstr{\"o}m} {et~al.}(2013){Wallstr{\"o}m}, {Biscaro}, {Salgado},
  {Black}, {Cherchneff}, {Muller}, {Bern{\'e}}, {Rho}, \&
  {Tielens}}]{wallstrom13}
{Wallstr{\"o}m}, S.~H.~J., {Biscaro}, C., {Salgado}, F., {et~al.} 2013, \aap,
  558, L2

\bibitem[{{Ziurys} \& {Turner}(1986)}]{ziurys86}
{Ziurys}, L.~M. \& {Turner}, B.~E. 1986, \apjl, 300, L19

\bibitem[{{Zubovas} \& {King}(2014)}]{zubovas14}
{Zubovas}, K. \& {King}, A.~R. 2014, \mnras, 439, 400

\end{thebibliography}

\end{document}